\documentclass[reprint,superscriptaddress,amsmath,amssymb,aps,pra,floatfix,]{revtex4-2} 
\usepackage{float}
\makeatletter
\let\newfloat\newfloat@ltx
\makeatother
\usepackage{subcaption} 
\usepackage{bm}
\usepackage{dcolumn}
\usepackage{bm}
\usepackage[english]{babel}
\usepackage[utf8]{inputenc}
\usepackage{amsthm}
\usepackage{mathtools}
\usepackage{physics}
\usepackage{algorithm}
\usepackage{algpseudocode}
\usepackage{amsmath}
\usepackage{xcolor}
\usepackage{graphicx}
\usepackage[left=23mm,right=13mm,top=35mm,columnsep=15pt]{geometry} 
\usepackage{adjustbox}
\usepackage{placeins}
\usepackage[T1]{fontenc}
\usepackage{lipsum}
\usepackage{csquotes}
\usepackage{float}
\usepackage{babel,blindtext}
\usepackage{hyperref}
\usepackage[normalem]{ulem}
\usepackage{quantikz}
\usepackage[title]{appendix}

\usepackage{footnote}
\usepackage{amssymb}
\usepackage{physics}
\usepackage{braket}
\usepackage{esvect}
\usepackage{caption}
\usepackage[color=yellow]{todonotes}
\begin{document}

\preprint{APS/123-QED} 

\title{The Harrow-Hassidim-Lloyd algorithm with qutrits} 

\author{Tushti Patel}
\email{tushti.oficial@gmail.com}
\affiliation{Centre for Quantum Engineering, Research and Education, TCG CREST, Kolkata 700091, India} 
\affiliation{Department of Physics, IIT Tirupati, Chindepalle, Andhra Pradesh 517619, India}
%\author{Akshaya Jayashankar}
%\affiliation{Centre for Quantum Engineering, Research and Education, TCG CREST, Kolkata 700091, India} 
\author{V. S. Prasannaa}
\email{srinivasaprasannaa@gmail.com}
\affiliation{Centre for Quantum Engineering, Research and Education, TCG CREST, Kolkata 700091, India} 
\affiliation{Academy of Scientific and Innovative Research (AcSIR), Ghaziabad 201002, India} 

\begin{abstract}
We extend the Harrow-Hassidim-Lloyd (HHL) algorithm, which is well-studied in the qubit framework, to its qutrit counterpart (which we call qutrit HHL, as opposed to qubit HHL, which is HHL using qubits), and develop a program for its implementation. We design Weyl-Heisenberg gadgets, the qutrit equivalents of Pauli gadgets, and come up with a practical implementation scheme for qutrit HHL. We test HHL with qutrits for simple matrices and verify the results against the expected outcomes. We apply the algorithm to quantum chemistry, and in particular, to the potential energy curve calculations of the model problem of the Hydrogen molecule in the split valence basis. We do so for two cases: 1-qutrit and 2-qutrit input states, where the latter makes use of our gadgets. We compare the number of qudits and the number of gates required between qubit and qutrit HHL implementations. In general, we find that for a fixed precision, the qutrit HHL circuit requires fewer number of qudits and comparable number of two-qudit gates than its qubit counterpart. 
\end{abstract} 

\maketitle 

\section{Introduction} 

The Harrow-Hassidim-Lloyd (HHL) algorithm \cite{HHL_original} is a quantum algorithm designed to solve systems of linear equations of the form $A\vec{x} = \vec{b}$, where $A$ is a matrix of dimension $N \cross N$, while $\vec{x}$ and $\vec{b}$ are vectors of dimension $N \cross 1$. HHL finds an approximation to $\ket{x} = A^{-1}\ket{b}$, where $\ket{b}$ is a suitably encoded version of $\vec{b}$. It is one of the few quantum algorithms that, in principle, promises an exponential advantage in system size. It has numerous applications in a wide variety of problems, such as portfolio optimization \cite{appli-portfolio_opt}, security against cyberattacks on a quantum computer \cite{appli-cyber_attack}, fluid dynamics \cite{appli-fluid_dynamics}, power flow \cite{appli-powerflow}, and solving the Poisson equation \cite{appli-poisson}. Recently, it was identified that HHL can be used to solve chemistry problems \cite{adaptHHL,psiHHL,appli-eigstate_prep} in addition to the existing variational quantum eigensolver (VQE) \cite{peruzzo2014variational} and quantum phase estimation (QPE) \cite{chem-qpe} methods. Quantum chemistry is identified to be one of the killer applications of quantum computing, and the analysis by the authors of References \cite{psiHHL,pbt} show that $\kappa$ and $s$, which refer to condition number of $A$ and the number of non-zero entries in the row with the maximum number of non-zero entries respectively, scale favorably (polylogarithmic in number of spin orbitals) within the limited molecules they consider. Furthermore, the system size, that is, the size of the A matrix itself, scales polynomially in the number of spin orbitals. Therefore, one can expect an advantage with HHL applied to chemistry. 

All of the above discussed works involve qubits. Despite the large volume of ongoing research in quantum algorithms and computing, there remains a relative paucity of literature on higher-dimensional qudits, and in particular, qutrits, where a qutrit has an additional quantum level relative to a qubit, but which is a natural extension to consider. Furthermore, in view of some of the recent advances in qutrit hardware \cite{qutrit-hardware,qutrit-hardware-gates,qutrit-benchmarking}, it is timely to explore the domain of quantum algorithms with qutrits. Therefore, given the potential of HHL for offering advantage across a wide variety of problems and the timely need to investigate quantum algorithms using qutrits, we extend the HHL algorithm to the qutrit framework (which we shall term as qutrit HHL hereafter). 

Recent works have identified that using qudits for quantum algorithms can have potential advantages over their qubit counterparts \cite{qudit-adv,qutrit-benchmarking}. Several algorithms like Deutsch-Josza, Bernstein-Vazirani, quantum Fourier transform (QFT), QPE, etc, have been discussed in the qudit framework \cite{qudit-algorithms}. In particular, literature on QFT dates back to as early as 2002 and 2007 \cite{qft_qudit1,qft_qudit2}. QPE also has been investigated extensively, and includes works like Reference \cite{qpe_qudits_first}, which develops the theoretical framework for QPE with qudits, and Reference \cite{qpe_qudits2}, in which the authors carry out some numerical simulations. Also, there have been recent works that have implemented the QFT and QPE algorithms on qudit hardware \cite{qft_qudits_expt,qpe_qudits_expt}. Since our goal is to develop qutrit HHL, we use the benefit of such developments from QPE towards our purpose. 

We discuss a scheme for the practical implementation of qutrit HHL, given a simple gate set comprising of 1- and 2-qutrit gates. To this end, we come up with qutrit equivalents of Pauli gadgets \cite{pauli_gadgets}, which we call Weyl-Heisenberg (WH) gadgets, to implement unitaries expressed as matrix exponentials. We also derive relations for qutrit operators, which include expressing any $3^n \times 3^n$ matrix (an $n-$ qutrit matrix) as a linear combination of tensor products of the nine $1-$ qutrit WH operators and an efficient circuit construction for the controlled-WH gadgets. 

Furthermore, given the potential of applying HHL to quantum chemistry, we consider it as a pilot application for qutrit HHL, and to that end, carry out calculations to generate the potential energy curve for the model system of $H_2$ in a split valence basis. This is followed by a comparative analysis of the costs involved between qubit and qutrit HHL calculations. To the best of our knowledge, one previous work proposes qutrit HHL, but only presents a preliminary circuit design \cite{old-qutrit-hhl}. 

The subsequent sections are organized as follows: Section \ref{sec:HHL} introduces the HHL algorithm for qubits. Following a primer on ternary logic and qutrit gates in Sections \ref{sec-ternary} and \ref{sec-gates} respectively, we move to our qutrit HHL implementation, which is discussed in Section \ref{sec:qutritHHL}. Next, in Section \ref{sec:decomposition}, we discuss in detail the decomposition of multi-qutrit unitaries to 1- and 2-qutrit gates, along with the step by step scheme to practically implement the qutrit HHL as a quantum circuit. We then provide a brief introduction to the quantum chemistry application, in which we solve for the correlation energy of a molecule using HHL (Section \ref{sec:hhllcc}). We then discuss results from our numerical simulations in Section \ref{sec:results}, which include toy matrices (Section \ref{subsec:toy}) and a quantum chemistry application (Section \ref{subsec:chem}). We then carry out a comparative analysis between qubits and qutrits in Section \ref{sec:scaling}, where we discuss estimates for the number of qudits required (Section \ref{subsec:numqubits}) and the number of 2-qudit gates incurred for HHL calculations in the qubit and qutrit cases (Section \ref{subsec:numgates}). Lastly, in Section \ref{sec:gates_phy_H} we discuss the gate counts for implementation of physical Hamiltonians, comparing both the HHL implementations. We finally conclude in Section \ref{sec:conclusion}. 

\section{The HHL algorithm}\label{sec:HHL} 

\begin{figure}[t]
    \centering
    \includegraphics[width=1.0\linewidth]{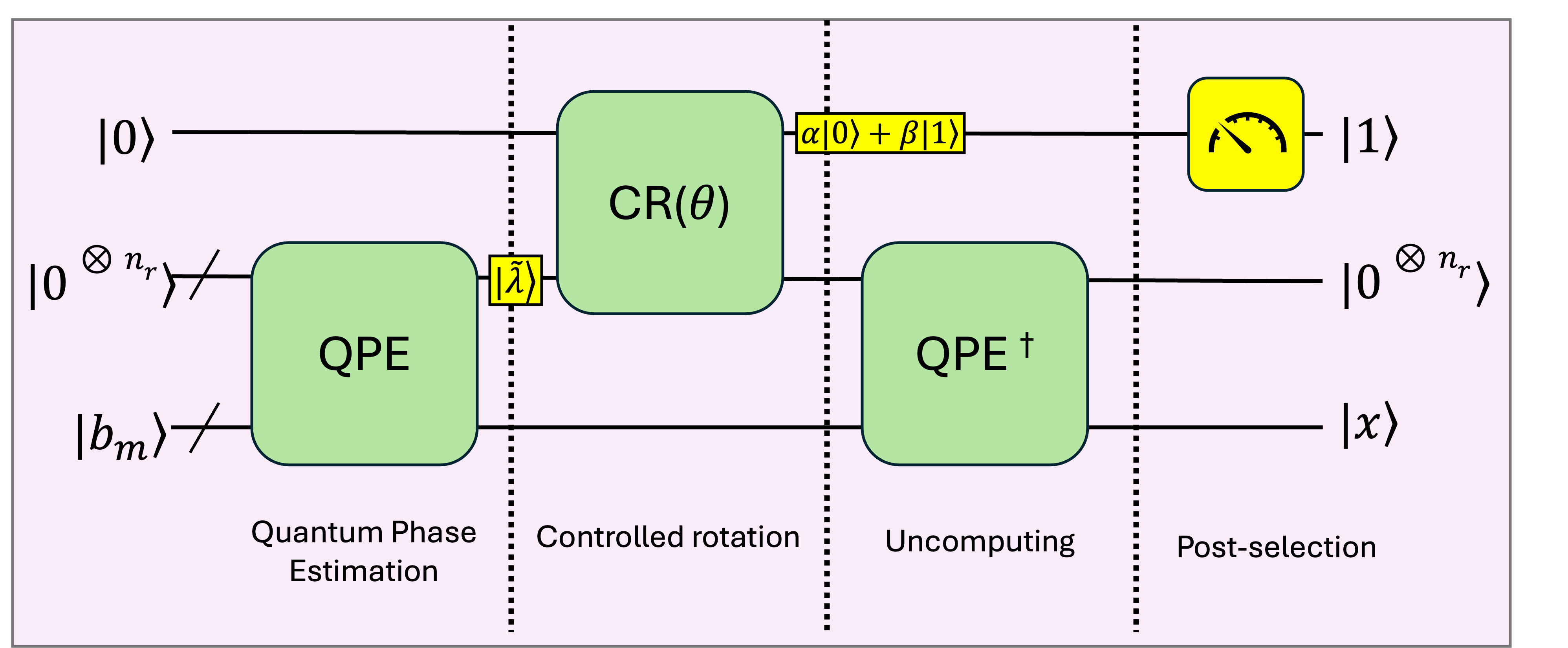}
    \caption{Schematic quantum circuit representation of the HHL algorithm. }
    \label{fig:hhl}
\end{figure}

The HHL algorithm is discussed in detail, for example, in Reference \cite{HHL_walkthru}, but here we provide only the relevant details that we would need for a subsequent qutrit extension of the algorithm. Figure \ref{fig:hhl} gives a schematic quantum circuit representation of the algorithm. The $N \cross 1 $ dimensional vector, $\vec{b}$, which is an input to the algorithm, is encoded in an $m$-dimensional normalized quantum state $|b\rangle_m$ in the state register via amplitude encoding such that $2^m = N$ for qubits. On the other hand, the matrix $A$ is used in the circuit as a part of a controlled-unitary of the form $C U$, where $U=e^{iAt}$ for an appropriately chosen scalar, $t$. As the figure shows, the HHL circuit broadly comprises of three parts - (a) QPE, (b) Controlled-rotation, and (c) Inverse of QPE. These steps are followed by measurement and post-selection. The above-mentioned controlled-unitaries, which are controlled on a set of $n_r$ clock register qubits, are used for estimating the phase. At the end of the QPE module, the scaled eigenvalues of $A$ are stored in the clock register with a precision of $n_r$ digits. The parametrized controlled-rotation module, controlled-$R(\theta)$, implements inversion of these eigenvalues using an ancillary qubit that we call the HHL ancilla for brevity. The inverse QPE module is then applied in order to disentangle the clock register from the rest of the system. Post-selection of $|1\rangle$ in the HHL ancilla results in a quantum state $|\Tilde{x}\rangle$ in the state register, which is proportional to the required $\vec{x}$. For any application, where some feature of $\vec{x}$ is to be evaluated, an additional module can be appended to the HHL circuit. For example, if the application of interest to us is quantum chemistry and we intend to compute molecular energies using HHL, then a controlled-swap test circuit module could be added at the end of the HHL circuit, as  discussed in Reference \cite{adaptHHL}. 

\section{HHL with qutrits} \label{sec:qutritHHL}

In this section, we discuss in detail the HHL algorithm using qutrits. 

\subsection{A brief introduction to ternary logic} \label{sec-ternary}

We begin our discussion with a concise introduction to ternary logic. Ternary logic is a multi-valued logic system that extends the binary system of logic by adding a third state `$2$' to \(\{0,1\}\). Thus, the computational basis for a ternary quantum state would be \(\{ |0\rangle,|1\rangle, |2\rangle \}\). A general 1-qutrit state can be defined as \(|\psi\rangle = \alpha_0 |0\rangle + \alpha_1 |1\rangle + \alpha_2 |2\rangle\) where \( |\alpha_0|^2 + |\alpha_1|^2 + |\alpha_2|^2 = 1\). Similarly, an $n$-qutrit state would have $3^n$ different possible basis states. In terms of the state vector notation, we can represent the 1-qutrit computational basis vectors as - 
\[
|0\rangle = 
\begin{pmatrix} 
    1 \\
    0 \\
    0
\end{pmatrix},
\quad
|1\rangle = 
\begin{pmatrix} 
    0 \\
    1 \\
    0
\end{pmatrix},
\quad \mathrm{and} \quad
|2\rangle = 
\begin{pmatrix} 
    0 \\
    0 \\
    1
\end{pmatrix}.
\]

\subsection{Basic gates for qutrits} \label{sec-gates} 

In the ternary framework, logical operators are extended to $3 \cross 3$ unitaries. In this case, the regular definitions of the qubit gates are not retained, and need to be modified. Just as the Pauli matrices $\sigma^x$, $\sigma^y$ and $\sigma^z$, along with identity form a complete basis for the space of operators acting on the Hilbert space for a qubit, we have the WH operators (Table \ref{tab:WH-ops}), which include identity, forming the basis for operators acting on qutrits \cite{WH_op_ref}. They are defined as - 
\begin{align*}
    X^a Z^b  \;\;\;\;\; \; \; \; \forall a,b \in \{ 0,1,2\}. 
\end{align*}

\begin{table*}[t]
\centering
\caption{Weyl--Heisenberg operators for a qutrit. Here, $\omega = e^{2\pi i/3}$.}
\renewcommand{\arraystretch}{1.2}
\begin{tabular}{c|ccccccccc}
\hline
$(a,b)$ 
& $(0,0)$ & $(1,0)$ & $(2,0)$ & $(0,1)$ & $(0,2)$ & $(1,1)$ & $(1,2)$ & $(2,1)$ & $(2,2)$ \\
\hline\hline
$W_{a,b}$ 
& $I$ & $X$ & $X^2$ & $Z$ & $Z^2$ & $XZ$ & $XZ^2$ & $X^2Z$ & $X^2Z^2$ \\
\hline
Matrix form
&
$\begin{pmatrix}
1&0&0\\0&1&0\\0&0&1
\end{pmatrix}$
&
$\begin{pmatrix}
0&0&1\\1&0&0\\0&1&0
\end{pmatrix}$
&
$\begin{pmatrix}
0&1&0\\0&0&1\\1&0&0
\end{pmatrix}$
&
$\begin{pmatrix}
1&0&0\\0&\omega&0\\0&0&\omega^2
\end{pmatrix}$
&
$\begin{pmatrix}
1&0&0\\0&\omega^2&0\\0&0&\omega
\end{pmatrix}$
&
$\begin{pmatrix}
0&0&\omega^2\\1&0&0\\0&\omega&0
\end{pmatrix}$
&
$\begin{pmatrix}
0&0&\omega\\1&0&0\\0&\omega^2&0
\end{pmatrix}$
&
$\begin{pmatrix}
0&\omega&0\\0&0&\omega^2\\1&0&0
\end{pmatrix}$
&
$\begin{pmatrix}
0&\omega^2&0\\0&0&\omega\\1&0&0
\end{pmatrix}$
\\
\hline
\end{tabular}
\label{tab:WH-ops}
\end{table*}

Some other gates that are used in this work are as follows:
\subsubsection{1-qutrit gates} 

\begin{enumerate}
    \item Hadamard gate (H): 
     \[
    H = \frac{1}{\sqrt{3}}
    \begin{pmatrix}
        1 & 1 & 1 \\
        1 & \omega & \omega^2 \\
        1 & \omega^2 & \omega
    \end{pmatrix} ,
    \]
    where $\omega = e^{2\pi i/3}$. In the outer product notation,
    \[
    H = \frac{1}{\sqrt{3}} \sum_{j,k \; \in \; \{0,1,2\}} \omega^{j\cdot k} |j\rangle \langle k|.
    \]

    The action of this gate can be described as follows - 
    \[
    \aligned
    H |0\rangle &= \frac{|0\rangle + |1\rangle + |2\rangle}{\sqrt{3}}  \\
    H |1\rangle &= \frac{|0\rangle + \omega |1\rangle + \omega^2 |2\rangle }{\sqrt{3}}  \\
    H |2\rangle &= \frac{|0\rangle + \omega^2 |1\rangle + \omega |2\rangle }{\sqrt{3}} .
    \endaligned
    \]
    \item S gate: 
     \[
    S = 
    \begin{pmatrix}
        1 & 0 & 0 \\
        0 & 1 & 0 \\
        0 & 0 & \omega
    \end{pmatrix} .
    \]
   
    The action of this gate can be described as follows - 
    \[
    \aligned
    S |0\rangle &= |0\rangle  \\
    S |1\rangle &= |1\rangle  \\
    S |2\rangle &= \omega |2\rangle .
    \endaligned
    \]
    
     \item Phase gate ($P_l$):
     \[
    P_l = 
    \begin{pmatrix}
        1 & 0 & 0 \\
        0 & e^{\frac{2 \pi i}{3^l}} & 0 \\
        0 & 0 & e^{\frac{4 \pi i}{3^l}}
    \end{pmatrix} .
    \]
    in the outer product notation,
    \[
    P_l = |0\rangle \langle 0| + e^{\frac{2 \pi i}{3^l}} |1\rangle \langle 1| + e^{\frac{4 \pi i}{3^l}} |2\rangle \langle 2 |.
    \]
    
    The action of this gate can be described as follows - 
    \[
    \aligned
    P_l |0\rangle &= |0\rangle \\
    P_l |1\rangle &= e^{\frac{2 \pi i}{3^l}} |1\rangle \\
    P_l |2\rangle &= e^{\frac{4 \pi i}{3^l}} |2\rangle .
    \endaligned
    \]
    \item Planar rotation gate ($R_{ij}(\theta)$):
    This gate performs rotation about the axis perpendicular to the $ij$ plane. For example, we may perform a rotation in the $|0\rangle - |1\rangle$ plane,
     \[
    R_{01}(\theta) = 
    \begin{pmatrix}
       \cos\!\left(\tfrac{\theta}{2}\right) & - \sin\!\left(\tfrac{\theta}{2}\right) & 0 \\[6pt]
        \sin\!\left(\tfrac{\theta}{2}\right) & \cos\!\left(\tfrac{\theta}{2}\right) & 0 \\[6pt]
        0 & 0 & 1
    \end{pmatrix} .
    \]
    
    The action of this gate can be described as follows - 
    \[
    \aligned
    R_{01}(\theta) |0\rangle &= \cos\left(\frac{\theta}{2}\right) |0\rangle + \sin \left( \frac{\theta}{2} \right) |1\rangle \\
    R_{01}(\theta) |1\rangle &=  - \sin \left(\frac{\theta}{2}\right) |0\rangle +  \cos \left(\frac{\theta}{2}\right) |1\rangle \\
    R_{01}(\theta) |2\rangle &= |2\rangle .
    \endaligned
    \]
\end{enumerate} 

\subsubsection{2-qutrit gates} 

\begin{enumerate}
    \item Controlled-increment ($CX$):
    The controlled-increment gate increments the state of the target qutrit depending on the control qutrit. This gate is also called the SUM gate as the operation is equivalent to a modulo 3 addition. The $CX$ gate is defined as -
    \[
    CX |j,k\rangle = |j,\; j + k (mod\; 3)\rangle,
    \]
    where $j$ is the control qutrit and $k$ is the target qutrit. In general, the action of the $CX$ gate can be described as -
    \[
    \aligned
    CX |0,k\rangle &= |0\rangle |k\rangle\\
    CX |1,k\rangle &= |1\rangle X|k\rangle\\
    CX |2,k\rangle &= |2\rangle X^2 |k\rangle,
    \endaligned
    \]
    where $|k\rangle $ is any 1-qutrit state, and the action of the $X$ gate is already defined in the list of 1-qutrit gates. 
    \item Controlled-$P_l$ gate:
    The action of the controlled-$P_l$ gate is described as - 
     \[
    \aligned
    CP_l |0,k\rangle &= |0\rangle |k\rangle\\
    CP_l |1,k\rangle &= |1\rangle P_l |k\rangle\\
    CP_l |2,k\rangle &= |2\rangle {P_l}^2 |k\rangle.
    \endaligned
    \]
    \item Generalized controlled-$R_{ij}(\theta)$:
    In this construction, $R_{ij}(\theta)$ gate acts on the target qutrit only if the state of the control qutrit is equal to $|m\rangle$, where $m$ is either 0, 1, or 2. 
\end{enumerate} 

\subsection{Qutrit HHL: working principles and quantum circuit} \label{sec:qutritHHL}

The HHL algorithm has been described briefly in Section \ref{sec:HHL}. In this section, we discuss our extension to the qutrit framework. The general problem statement remains the same, that is, solving a system of linear equations of the form $A \vec{x} = \vec{b}$. The $\vec{b}$ of dimension $N \cross 1$ is encoded in a state register containing $m$ qutrits using amplitude encoding such that $3^m = N$. In addition to the state register, we have a clock register comprising of $n_r$ qutrits. These are useful for capturing the scaled eigenvalues via the QPE module. Lastly, we have a single qutrit ancilla, also known as the HHL ancilla, which will be used for eigenvalue inversion through the controlled-rotation module. 

\begin{figure}
    \centering
    \includegraphics[width=1.0\linewidth]{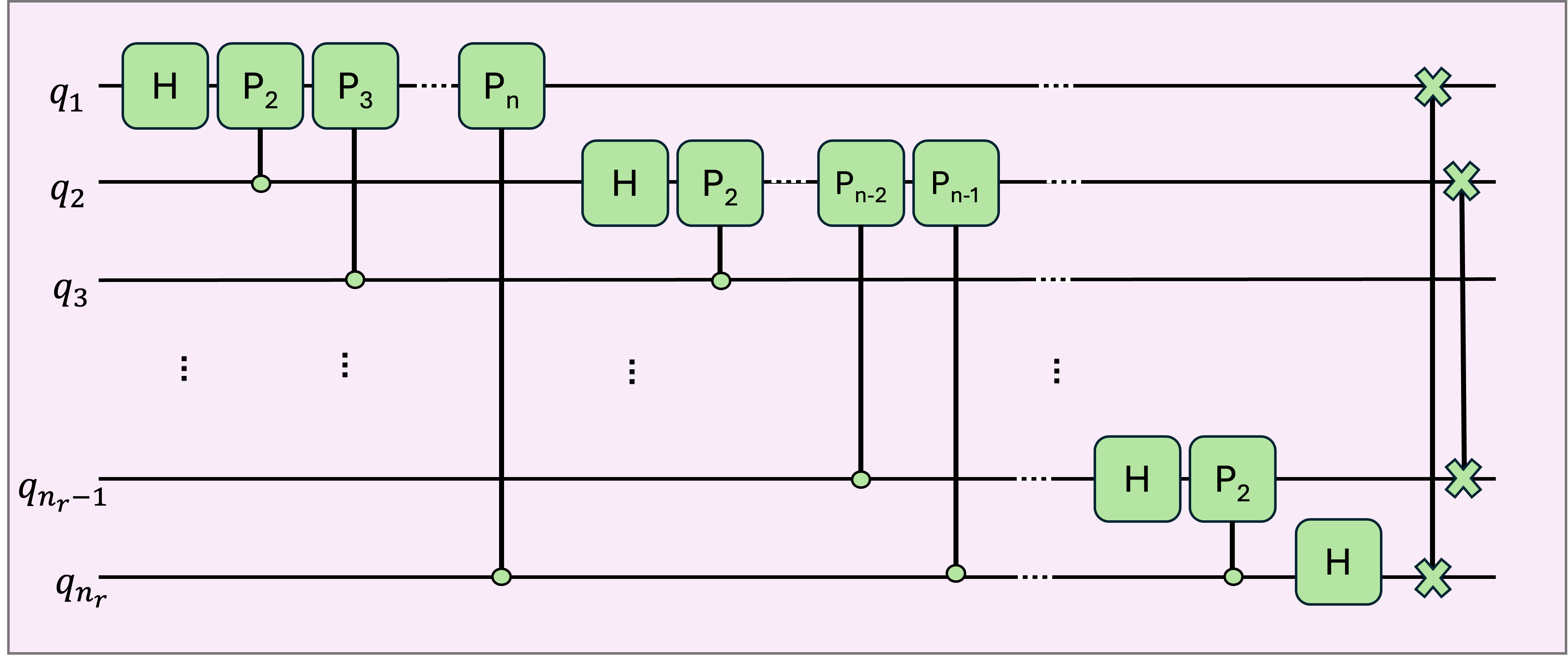}
    \caption{Quantum Fourier transform circuit for qutrits. }
    \label{fig:qutrit-QFT}
\end{figure}

We begin by discussing the QFT module, since its inverse (IQFT) occurs in the QPE module. Figure \ref{fig:qutrit-QFT} gives the circuit for QFT. Although the circuit is similar to that for qubits, the gate definitions are modified in the qutrit framework. We define the generalized $P_l$ gate for qutrits as specified in Section \ref{sec-gates}. Implementing the QFT circuit on a general n-qutrit state $|j_1, j_2, \hdots j_n \rangle$, we obtain -
\[
QFT \; |j_1, j_2, \hdots j_n \rangle = \frac{1}{3^{n/2}} \sum_{k=0}^{3^n-1} e^{\frac{2 \pi i (j.k)}{3^n}} |k_1, k_2, \hdots k_n \rangle.
\]

\begin{figure}
    \centering
    \includegraphics[width=1.0\linewidth]{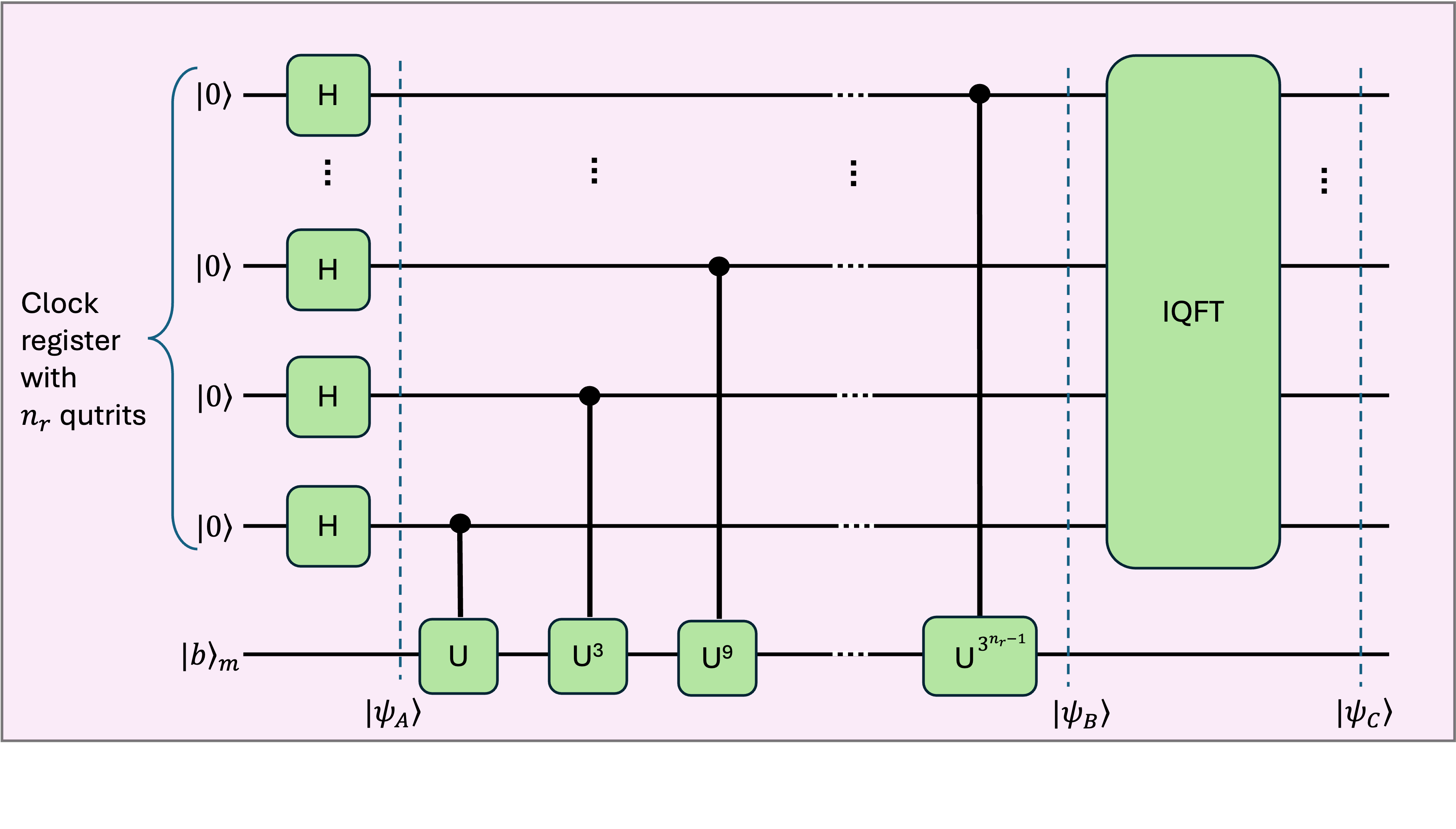}
    \caption{Quantum phase estimation circuit for qutrits. }
    \label{fig:qutrit-QPE}
\end{figure} 

The circuit for the QPE module is given in Figure \ref{fig:qutrit-QPE}. QPE consists of mainly controlled-unitaries, with the unitary matrix of the form $U = e^{iAt}$. Here, $A$ is the Hermitian matrix that occurs in the system of linear equations to be solved, and $t$ is a problem-dependent constant such that $U(\sum_{i=0}^{3^m-1} b_i|u_i\rangle ) = \sum_{i=0}^{3^m-1} b_i e^{2 \pi i \phi_i} |u_i\rangle$ where $\phi_i$ $\in [0,1)$ are the scaled eigenvalues, and are related to the eigenvalues of $A$, $\{\lambda_i\}$, as $\lambda_i \;t = 2 \pi \phi_i , \forall i$. The clock registers are initially all in state $|0\rangle$. First, all the clock registers are acted upon by the ternary Hadamard gates resulting in -

\begin{align*}
    |\psi_A\rangle =  \left(\frac{|0\rangle + |1\rangle + |2\rangle}{\sqrt{3}}\right)^{\otimes n_r} \otimes |b\rangle_m. 
\end{align*}
Upon the application of the series of controlled-unitaries shown in Figure \ref{fig:qutrit-QPE}, we obtain 
\begin{align*}
    |\psi_B\rangle = \sum_{i=0}^{3^m-1} b_i & \left(\frac{|0\rangle + e^{3^{n_r-1}\;2 \pi i \phi_i}|1\rangle +  e^{3^{n_r-1} \; 4 \pi i \phi_i}|2\rangle }{\sqrt{3}}\right) \otimes\\ & \left(\frac{|0\rangle + e^{3^{n_r-2}\;2 \pi i \phi_i}|1\rangle +  e^{3^{n_r-2} \; 4 \pi i \phi_i}|2\rangle }{\sqrt{3}}\right) \otimes  \\
    \hdots &\otimes \left(\frac{|0\rangle + e^{2 \pi i \phi_i}|1\rangle + e^{4 \pi i \phi_i}|2\rangle }{\sqrt{3}}\right) \otimes |u_i\rangle_m .
\end{align*}

Substituting $\phi_i$ in the form $ \Tilde{\phi_i} = 0.j_1j_2\hdots j_{n_r}^{(i)}$, we obtain an expression equivalent to that of QFT, and hence the inverse QFT module takes the scaled eigenvalues of $A$ from the phase to the qutrit state. The precision is, of course, limited by the number of clock register qutrits used and the resultant state after the QPE module $|\psi_C\rangle = \sum_{i=0}^{3^m-1} b_i \Tilde{|\phi_i \rangle} \otimes |u_i\rangle_m$. Reference \cite{qudit-algorithms} briefly describes the QFT and QPE modules for higher-dimensional qudits, in general. We have worked them out in detail for the specific case of qutrits. 

\begin{figure}
    \centering
    \includegraphics[width=0.80\linewidth]{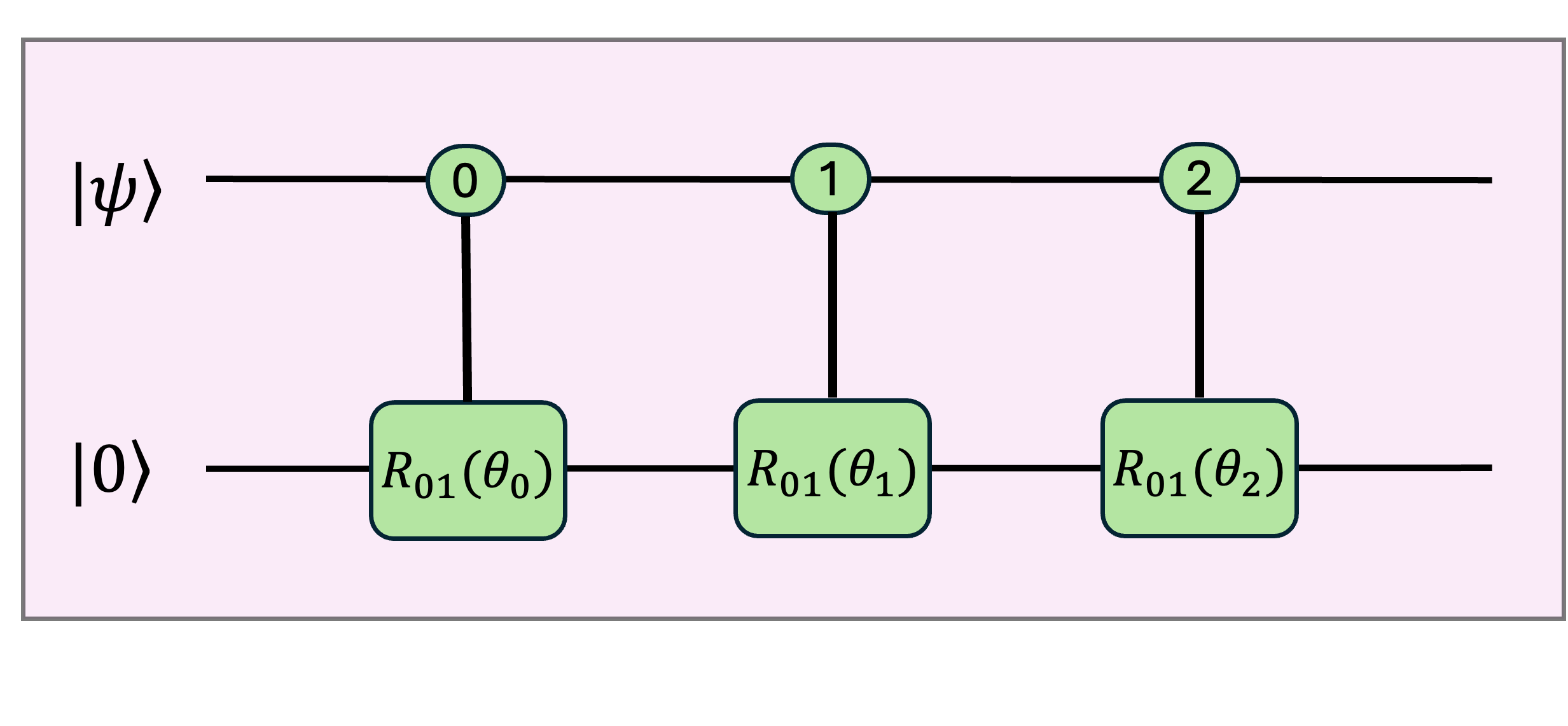}
    \caption{Uniformly controlled-rotation circuit for 1 control qutrit. }
    \label{fig:qutrit-UCR}
\end{figure}

Next, we have the parametrized controlled-rotation module, which is used for eigenvalue inversion. Rotations in general could be quite complex in a qutrit framework, and for the purposes of our HHL algorithm, we choose to use planar rotations, that is, a rotation of any two of the three orthogonal qutrit basis states. Without loss of generality, we consider $|0\rangle$ and $|1\rangle$ for our rotations (any other combination can be used equivalently). The rotation gate as well as the controlled-rotation gates have been described in detail in Section \ref{sec-gates}. The simple example circuit shown in Figure \ref{fig:qutrit-UCR} serves as an illustration of the workings of uniformly controlled-rotations (UCR). We assume that the control qutrit is in a general state $|\psi\rangle = \alpha_0 |0\rangle + \alpha_1 |1\rangle + \alpha_2 |2\rangle $. The action of the three UCR gates can be described in the outer product notation as $|0\rangle \langle 0| \otimes R_{01}(\theta_0) + |1\rangle\langle 1| \otimes R_{01}(\theta_1) + |2\rangle\langle 2| \otimes R_{01}(\theta_2) $. Therefore, 
\begin{align*}
    UCR (|\psi \rangle &\otimes |0\rangle) = \sum_{i=0}^2 \alpha_i |i\rangle \\ &\otimes \left( \sqrt{1 - \sin^2 \left( \frac{\theta_i}{2} \right)} |0\rangle + \sin \left(\frac{\theta_i}{2}\right) |1\rangle \right).
\end{align*}

Substituting for $\theta_i = 2 \; \sin^{-1}\left( \frac{\mathcal{C}}{\Tilde{\phi_i }}\right)$, where $\mathcal{C}$ is a suitably chosen constant which is generally taken as the minimum scaled eigenvalue of $A$, we get
\[
UCR (|\psi \rangle \otimes |0\rangle) = \sum_{i=0}^2 \alpha_i |i\rangle \otimes \left(\sqrt{1 - \left(\frac{\mathcal{C}}{\Tilde{\phi_i }}\right)^2} |0\rangle + \frac{\mathcal{C}}{\Tilde{\phi_i }} |1\rangle \right).
\]
Thus, the coefficient of $|1\rangle$ is the required inverse of the scaled eigenvalues. The same logic can be extended to multiple control qutrits with multi-controlled rotation gates. The  clock register acts as the set of control qutrits. 

Lastly, the inverse QPE module and measurement of the HHL ancilla qutrit followed by post-selection of $|1\rangle$ are implemented to obtain $|\Tilde{x}\rangle$ in the state register, which is proportional to the required $\vec{x}$ up to known constants. 

\section{Towards a practical implementation of the qutrit HHL} \label{sec:decomposition} 

A very important part of a practical implementation of the qutrit HHL algorithm is the gadgets to use in decomposing the controlled-unitary gate that occurs in QPE. In fact, although our interest is in qutrit HHL, deriving such gadgets is useful for standalone Hamiltonian simulation and QPE calculations, as well as for other algorithms that use QPE. We feed in $A$ into the controlled-unitary, where $U = \;e^{iAt}$, and after expressing $A$ in the WH operator basis (Section \ref{basis-expn}) and Trotterizing (first order and first step), we arrive at a sequence of matrix exponentials each containing a WH operator in its exponent. Just as one uses Pauli gadgets to decompose each such exponential into a compact set of 1- and 2- qubit gates ($\sim n$ 2-qubit gates for an $n-$qubit circuit), we show in Section \ref{ckt-wh} that one can use WH gadgets for the qutrit counterpart. Different schemes are found in literature for decomposition of higher dimensional unitaries \cite{shende_decompo,jiang_qutrit_decompo}. Recently, there has been a work on the topic of decomposition for multi-qutrit gates \cite{decomposition_WHgadgets}, where they extend the stair-case structure of controlled-gates that occur in qubits to the qutrit framework, for the case of $Z$ and $Z^2$ as well as their tensor products occurring as the exponent (see Equation (12) and Figure 4 of Reference \cite{decomposition_WHgadgets}). However, they do not delve into the details of unitaries that can transform the other WH operators in terms of $Z$ and $Z^2$. We explicitly derive those transformations in Equations (\ref{eq:XZ_decompo}) through (\ref{eq:X2Z_decompo}). Figure \ref{fig:gadgets} shows circuit diagrams for all possible WH gadgets for a 2-qutrit decomposition.
Similarly, taking inspiration from the results of Reference \cite{blunt_controlled_pauligadget} for qubits on decomposing a controlled-unitary, such that the unitary is generated using a Pauli string, we introduce a qutrit equivalent for decomposing a controlled unitary (see Equations (\ref{eq:G_def}) through (\ref{eq:bluntforqutrit})). Following is an outline of the stepwise implementation of qutrit HHL along with derivations wherever needed. 

\subsection{Expressing the $A$ matrix in terms of WH operators}\label{basis-expn} 

We can express any matrix of the form $2^n \times 2^n$ as a linear combination of tensor products of Pauli matrices. The coefficients of the expansion can be obtained using the relation \cite{nielsenChuang} - 

\begin{align} 
A &= \sum_{k}
c_k\, P_k, \\
c_k &= \frac{1}{2^n}\operatorname{Tr}\!\left(P_k^{\dagger} A\right).
\end{align}
Extending this relation to qutrits using the WH operators, for any matrix of dimension $3^n \times 3^n$ we get - 

\begin{align}
A &= \sum_{a,b}
( \, c_{a,b}\, W_{a,b}, + c_{a,b}^* \, W_{a,b}^\dagger \,)\label{eq:A_expansion} \\
c_{a,b} &= \frac{1}  {3^n}\operatorname{Tr}\!\left(W_{a,b}^{\dagger} A\right). \label{eq:A_expa_coeff}
\end{align}

In the above equations, each $W_{a,b}$ is a tensor product of $n$ WH matrices shown in Table \ref{tab:WH-ops}, and thus the composite index pair $a,b$ runs over $3^{2n}$ values of the individual $a,b$ values of each qutrit. For a Hermitian $A$ matrix, $(A=A^\dagger)$, and hence 

\begin{align*}
    c_{a,b}^* &=\frac{1}{3^n}\operatorname{Tr}( (W_{a,b}^\dagger A)^\dagger) \\
&=\frac{1}{3^n}\operatorname{Tr}(AW_{a,b}) \\
&=\frac{1}{3^n}\operatorname{Tr}(W_{a,b}A),
\end{align*}

Now, using Trotterization \cite{trotter} (first order and Trotter step set to 1), we can approximate -
\begin{align}
    e^{iAt} &=  e^{it\sum_{a,b}
(c_{a,b}\, W_{a,b}\, + c_{a,b}^* \, W_{a,b}^\dagger \,)} \nonumber \\  &\approx \prod_{a,b}e^{it\,
( \, c_{a,b}\, W_{a,b}\, + c_{a,b}^* \, W_{a,b}^\dagger \,)} .  \label{eq:trotter}
\end{align}
The next task is to establish that each piece in the above summation is unitary. The authors of Reference \cite{decomposition_WHgadgets} show that as long as the exponent (barring $it$) is $cZ^n \otimes Z^n \otimes \cdots \otimes Z^n + h.c.$, a unitary circuit exists for the expression (see Equation (35) of their Appendix). If we can show that for the cases involving other $W_{a,b}$ operators, that is, suitable rotations exist such that these operators can be rotated to Equation (35) of their work, we have established the circuits for all possible terms in Equation (\ref{eq:trotter}). We undertake this task in the succeeding sub-section. 

\subsection{Circuit implementation of the unitary $e^{iAt}$ in terms of WH gadgets}\label{ckt-wh}

Let us begin by deriving the 2-qutrit gadget for the simplest case of $W_{a,b}=Z \otimes Z$. In the rest of this sub-section where we discuss the case of two qutrits, the indices $a,b$ run over only the first, since we fix the second qutrit to $Z$ without loss of generality. Using the definition of $CX$ gate as -
\begin{align*}
CX
&=
\sum_{j=0}^{2} |j\rangle\!\langle j| \otimes X^{j}, 
\end{align*}
and certain identities like $ZX = \omega XZ$, we can show that -
\begin{align*}
CX^\dagger\,(I \otimes Z)\,CX
&=
\sum_{j=0}^{2} |j\rangle\!\langle j| \otimes X^{-j} Z X^{j} \\
&=
\sum_{j=0}^{2} \omega^{j} |j\rangle\!\langle j| \otimes Z \\
&=
Z \otimes Z.
\end{align*}
This can be extended to $Z^2 $ which is equal to $ Z^\dagger$ via -
\begin{align}
Z^{\eta} \otimes Z
&=
CX^{\dagger \eta} (I \otimes Z) CX^{\eta}, \; \; \; \eta \in \{1,2\} .
\end{align}

\begin{figure*}[htbp]
\centering

\begin{subfigure}{0.45\textwidth}
  \centering
  \includegraphics[width=\linewidth]{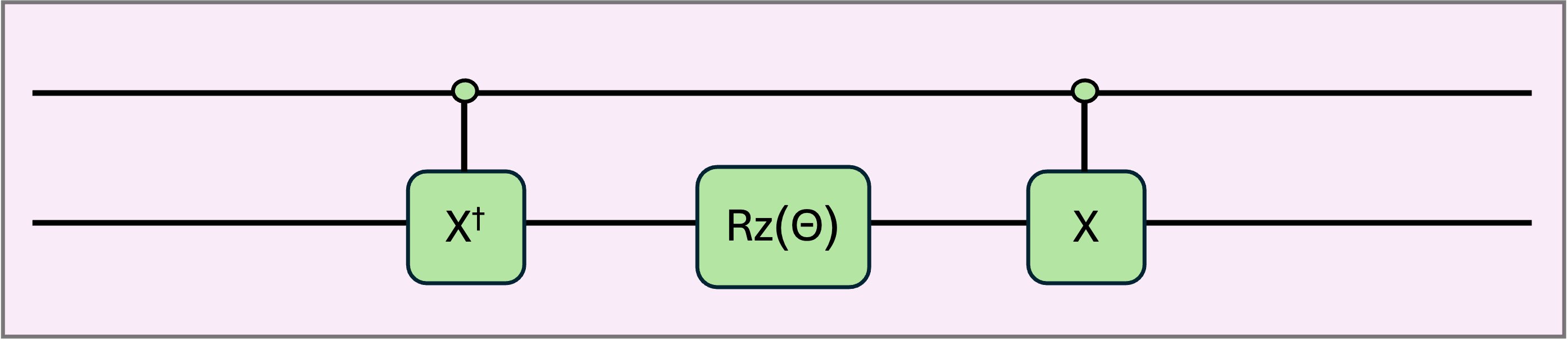}
  \caption{The WH gadget for $e^{i \frac{\Theta}{2} (c_{0,1}Z \otimes Z + h.c.)}$.}
\end{subfigure}\hfill
\begin{subfigure}{0.45\textwidth}
  \centering
  \includegraphics[width=\linewidth]{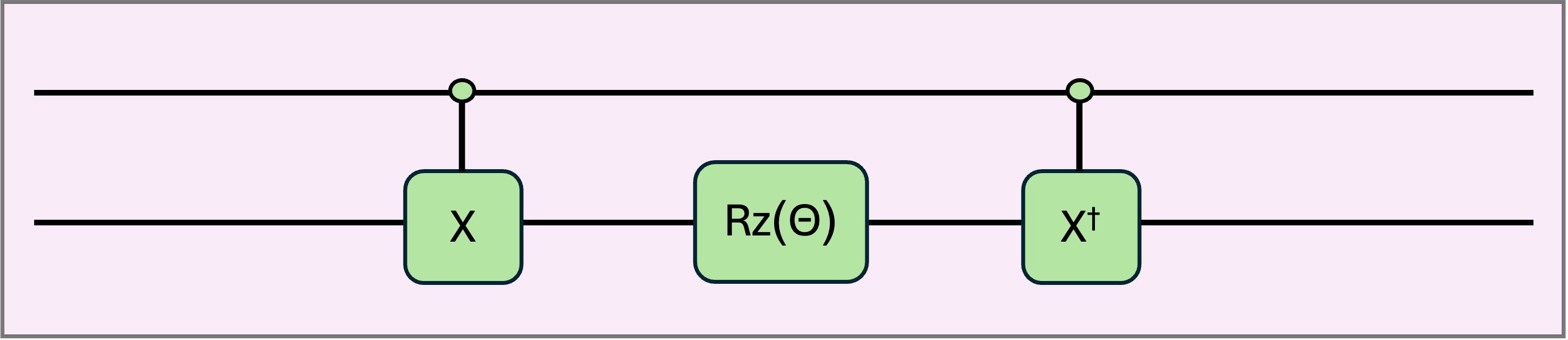}
  \caption{The WH gadget for $e^{i \frac{\Theta}{2} (c_{0,2}Z^2 \otimes Z + h.c.)}$.}
\end{subfigure}

\vspace{0.3cm}

\begin{subfigure}{0.45\textwidth}
  \centering
  \includegraphics[width=\linewidth]{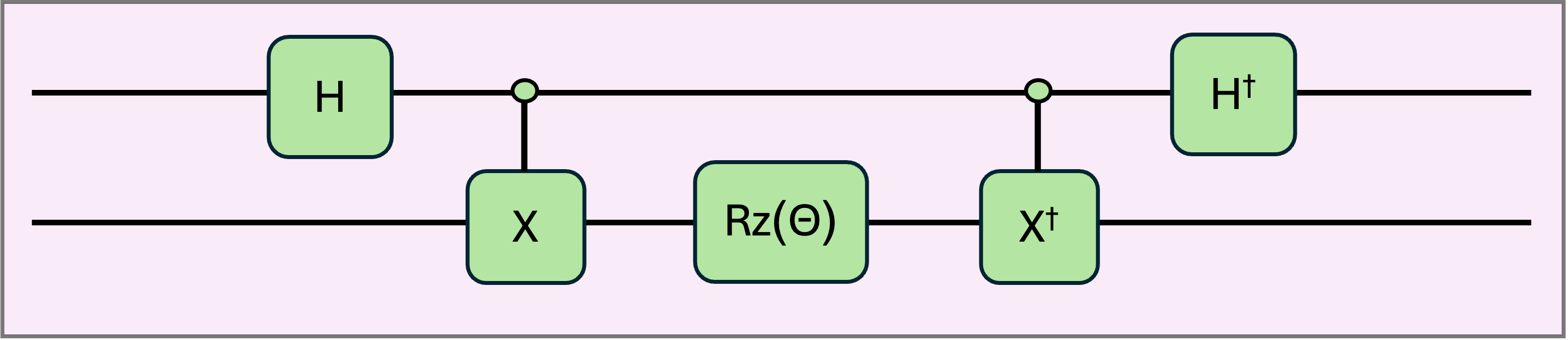}
  \caption{The WH gadget for $e^{i \frac{\Theta}{2} (c_{1,0}X \otimes Z + h.c.)}$.}
\end{subfigure}\hfill
\begin{subfigure}{0.45\textwidth}
  \centering
  \includegraphics[width=\linewidth]{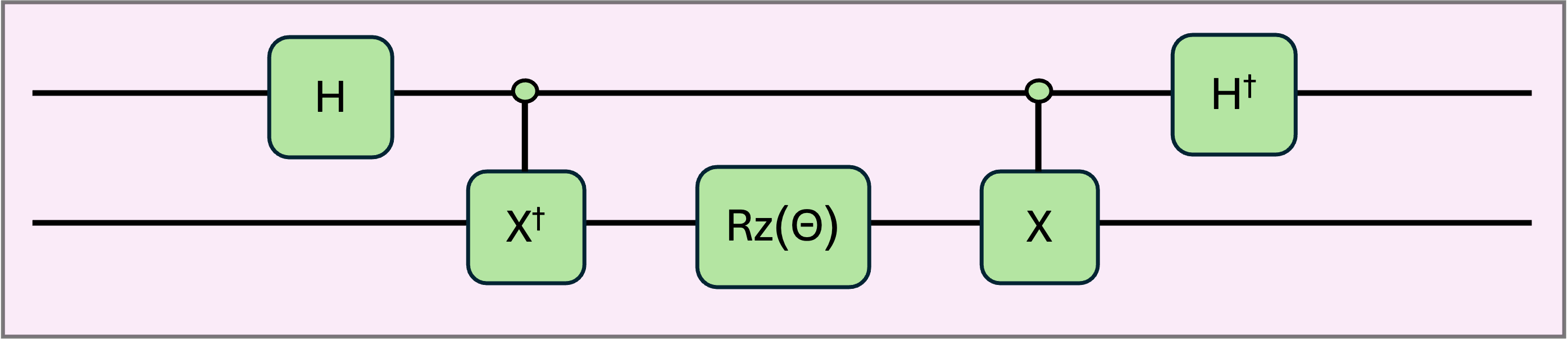}
  \caption{The WH gadget for $e^{i \frac{\Theta}{2} (c_{2,0}X^2 \otimes Z + h.c.)}$.}
\end{subfigure}

\vspace{0.3cm}

\begin{subfigure}{0.45\textwidth}
  \centering
  \includegraphics[width=\linewidth]{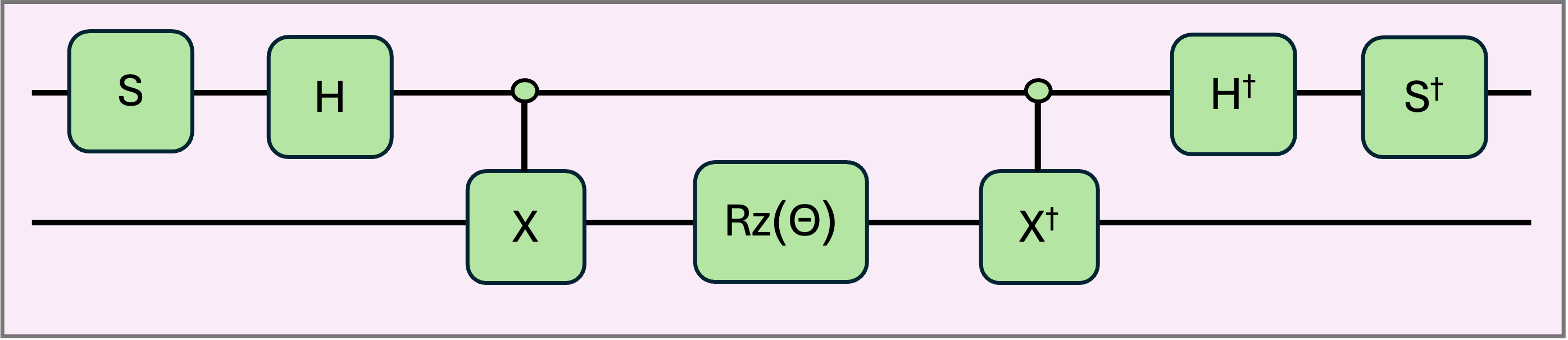}
 \caption{The WH gadget for $e^{i \frac{\Theta}{2} (c_{1,1}XZ \otimes Z + h.c.)}$.}
\end{subfigure}\hfill
\begin{subfigure}{0.45\textwidth}
  \centering
  \includegraphics[width=\linewidth]{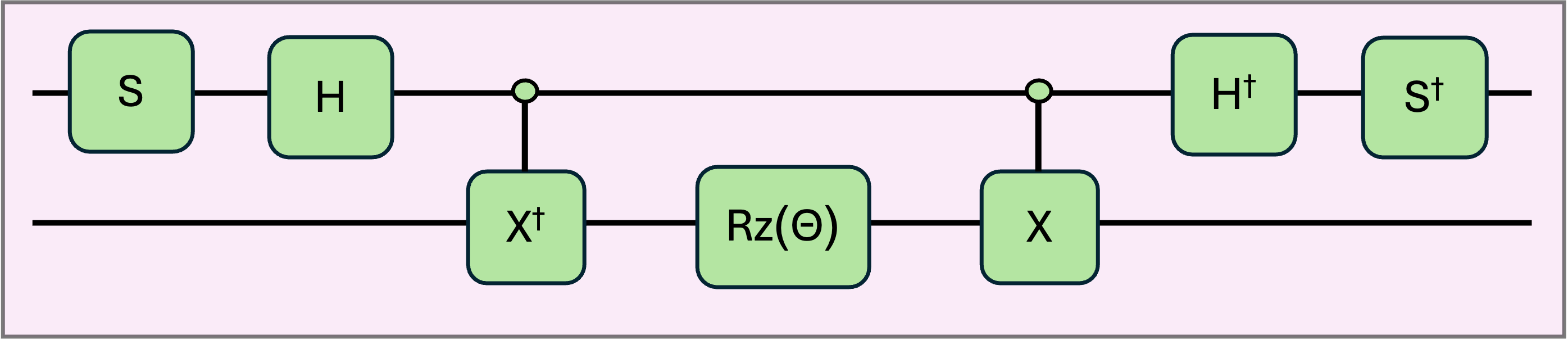}
  \caption{The WH gadget for $e^{i \frac{\Theta}{2} (c_{2,2}X^2Z^2 \otimes Z + h.c.)}$.}
\end{subfigure}

\vspace{0.3cm}

\begin{subfigure}{0.45\textwidth}
  \centering
  \includegraphics[width=\linewidth]{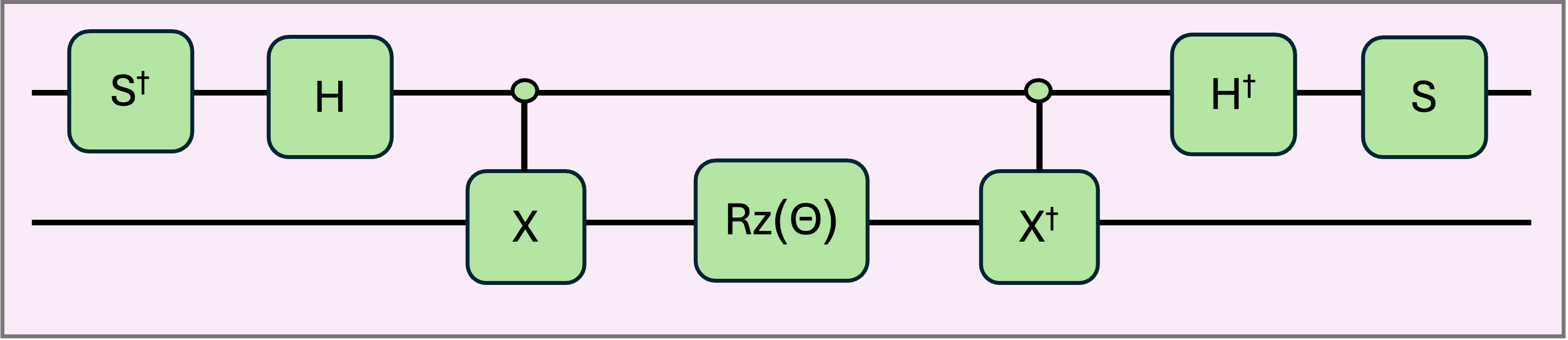}
  \caption{The WH gadget for $e^{i \frac{\Theta}{2} (c_{1,2}XZ^2 \otimes Z + h.c.)}$.}
\end{subfigure}\hfill
\begin{subfigure}{0.45\textwidth}
  \centering
  \includegraphics[width=\linewidth]{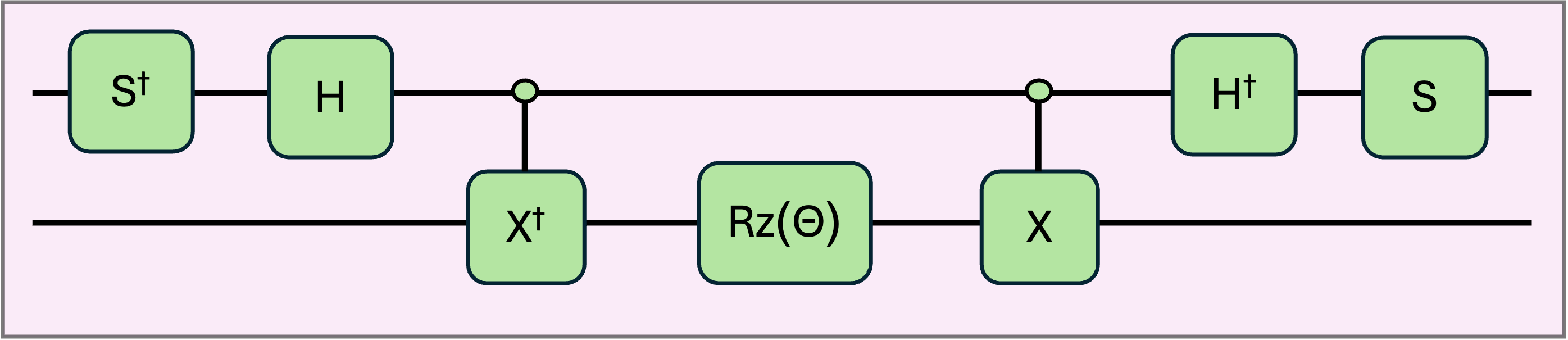}
  \caption{The WH gadget for $e^{i \frac{\Theta}{2} (c_{2,1}X^2Z \otimes Z + h.c.)}$.}
\end{subfigure}

\caption{WH gadgets for all possible WH operators (excluding the trivial operator - Identity) acting on the first qutrit. Equivalent transformations can be applied to any other qutrit, for the respective WH operators acting on that qutrit. Here, $R_Z (\Theta) = R_{01}(Re(c_{a,b})\Theta) \; R_{02}(Re(c_{a,b})\Theta) \; R_{12}(\sqrt3 \; Im(c_{a,b})\Theta) $ for the respective $(a,b)$.} 
\label{fig:gadgets}
\end{figure*}

\begin{figure}[h]
    \centering
    \includegraphics[width=0.85\linewidth]{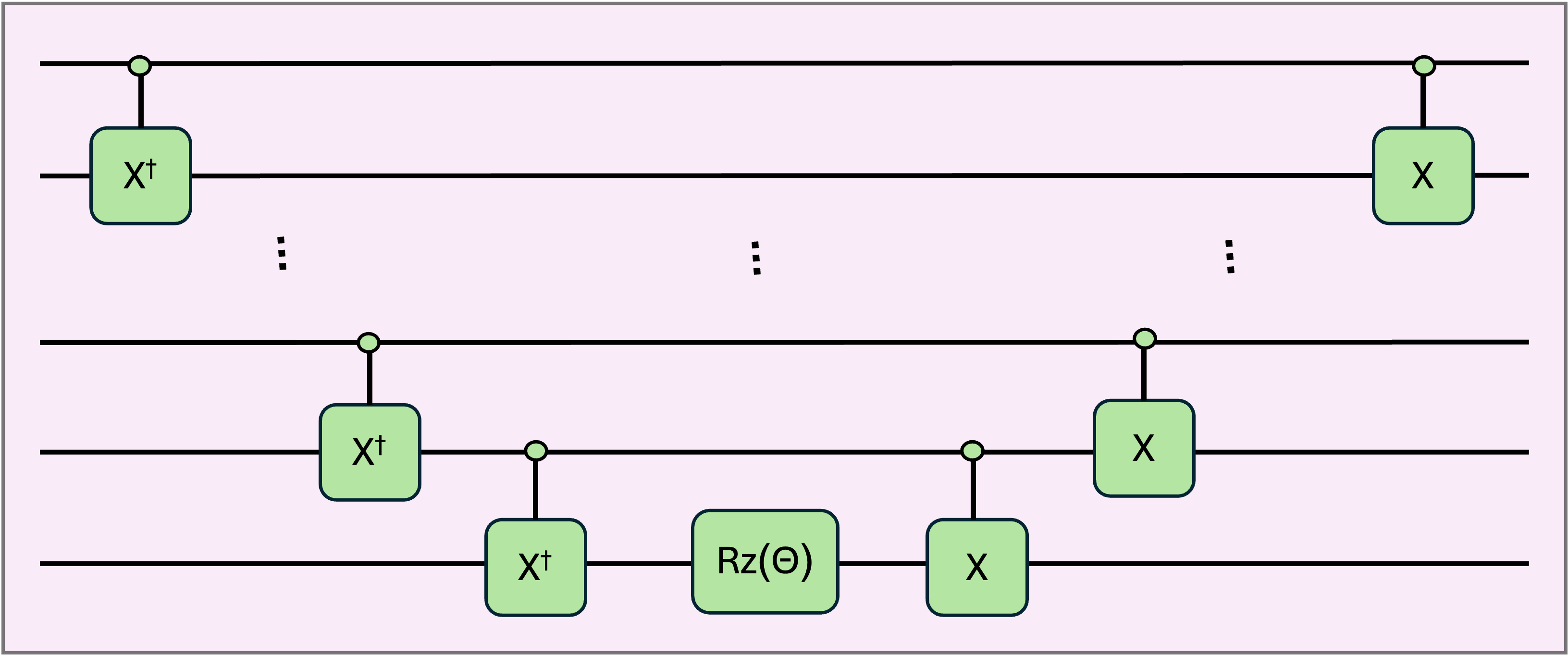}
    \caption{The WH gadget for $e^{i \frac{\Theta}{2} (c_{0,1}Z \otimes Z \otimes \hdots \otimes Z + h.c.)}$. Here, $R_Z (\Theta) = R_{01}(Re(c_{0,1})\Theta) \; R_{02}(Re(c_{0,1})\Theta) \; R_{12}(\sqrt3 \; Im(c_{0,1})\Theta) $.} 
    \label{fig:WH_staircase}
\end{figure}

Now, using the well-known relation-
\begin{align}
e^{i\Theta U P U^\dagger}
&=
U e^{i\Theta P} U^\dagger, \label{eq:exp_unitary}
\end{align}
which is valid for any unitary \(U\) and operator \(P\), we can show for any linear combination of $Z $ and $Z^2 $ that -

\begin{align}
e^{\frac{i\Theta}{2}\left(c_{0,\eta}Z^\eta \otimes Z + \text{h.c.}\right)}
&=
CX^{\dagger \eta}\;
e^{\frac{i\Theta}{2}\left(c_{0,\eta}I \otimes Z + \text{h.c.}\right)}
\;CX^{\eta}. \label{eq:ZtensorZ}
\end{align}
We have in fact re-derived Equation (35) of Reference \cite{decomposition_WHgadgets}. We have also made the substitution $t \rightarrow \Theta/2$. The $e^{\frac{i\Theta}{2}\left(c_{0,\eta}I \otimes Z + \text{h.c.}\right)}$ can be implemented as a parametrized rotation about the z-axis for the second qutrit using the following rotation module: $R_Z (\Theta) = R_{01}(Re(c_{0,\eta})\Theta) \; R_{02}(Re(c_{0,\eta})\Theta) \; R_{12}(\sqrt3 \; Im(c_{0,\eta})\Theta) $. The corresponding circuit is shown in Figure \ref{fig:gadgets}(a). In a similar way, the above logic can be extended to multiple qutrits via the staircase structure consisting of ${CX}^{\dagger \eta}$s to the left and ${CX}^{\eta}$s to the right, as shown in Figure \ref{fig:WH_staircase}. 

We can further show that all the possible WH operators can be transformed to $Z^2$ or $Z$ via unitary operations. We derive these transformations for all the WH operators given in Table \ref{tab:WH-ops}. Apart from the $Z$ and $Z^2$ operators, that can be directly implemented using Equation (\ref{eq:ZtensorZ}) we can show that $X = H Z^2 H^\dagger$ and $X^2 = H Z H^\dagger$. The $XZ$ operator can be expressed as $SXS^\dagger$, where $X$ can be further broken down in terms of $H$ and $H^\dagger$, resulting in -
\begin{align}
    XZ &= S(HZ^2H^\dagger)S^\dagger. \label{eq:XZ_decompo}
\end{align}
Now, on squaring Equation (\ref{eq:XZ_decompo}), we get -
\begin{align*}
    (XZ)^2 &= (SHZ^2H^\dagger S^\dagger)^2 \\
    XZXZ  &= (SHZ^2H^\dagger S^\dagger)(SHZ^2H^\dagger S^\dagger) \\
    \omega X^2 Z^2 &= SHZ^4H^\dagger S^\dagger \\
    X^2 Z^2 &= SHZH^\dagger S^\dagger, 
\end{align*}
we shall absorb the global phase factor $\omega$ in the coefficient in the exponent. We can also show that $Z = SZS^\dagger$, and using it along with Equation (\ref{eq:XZ_decompo}), we  obtain -
\begin{align*}
    XZ^2 &= (XZ)Z \\
         &= (SX S^\dagger)(SZS^\dagger) \\
         &= SXZS^\dagger \\
         &= S(S(HZ^2H^\dagger)S^\dagger)S^\dagger ,
\end{align*}
which can be further simplified as -
\begin{align}
    XZ^2 &= S^\dagger HZ^2H^\dagger S . \label{eq:XZ2_decompo}
\end{align}
The last WH operator that remains is $X^2Z$, which can be obtained by squaring Equation (\ref{eq:XZ2_decompo}) as follows -
\begin{align*}
(XZ^{2})^{2}
&= X Z^{2} X Z^{2} \\
&= \omega^{2} X^{2} Z^{4} \\
&= \omega^{2} X^{2} Z .
\end{align*}
We can equate this to the right hand side of squared Equation (\ref{eq:XZ2_decompo}) and again the global phase $\omega^2$ can be absorbed in the coefficient. Therefore,
\begin{align}
X^{2} Z
&=  \left( S^{2} H Z^{2} H^{\dagger} S^{\dagger} \right)^{2} \label{eq:X2Z_decompo} \\ \nonumber
&= S^{\dagger} H Z H^{\dagger} S .
\end{align}

When any of the above WH operators occur on any one or both of the qutrits, they can be pulled outside the $CX^{\eta}$ and $CX^{\dagger \eta}$ occurring in Equation (\ref{eq:ZtensorZ}) using the relation in Equation (\ref{eq:exp_unitary}). We show a simple demonstration of how the above-mentioned transformations translate into circuits for the case of $X$ occurring on the first qubit.
\begin{align*}
e^{i\Theta (c_{1,0}X \otimes Z \; + \; h.c.)}
&= e^{i\Theta (c_{1,0}X \otimes Z \; + \; c_{1,0}^* X^\dagger \otimes Z^\dagger)}\\
&=
e^{i\Theta (c_{1,0}H Z^2 H^\dagger \otimes Z \; + \; c_{1,0}^* H Z H^\dagger \otimes Z^\dagger)} 
\end{align*}
Since $Z$ and $Z^2$ commute,
\begin{align*}
e^{i\Theta (c_{1,0}X \otimes Z \; + \; h.c.)}
&=
e^{i\Theta \bigl( (H \otimes I)(c_{1,0}Z^2 \otimes Z \; + \; c_{1,0}^* Z \otimes Z^2 )(H^\dagger \otimes I) \bigr)} \\
&=
(H \otimes I)\
e^{i\Theta (c_{1,0}Z^2 \otimes Z \;+ \; h.c.)}\,
(H^\dagger \otimes I).
\end{align*}
We can further expand $e^{i\Theta (c_{1,0}Z^2 \otimes Z + h.c.)}$ using Equation (\ref{eq:ZtensorZ}). Circuit diagrams for the WH gadgets with all possible WH operators acting on the first qutrit are illustrated in Figure \ref{fig:gadgets}. 

In this way, one can express $U = e^{iAt}$ using Trotterization (Equation (\ref{eq:trotter})), with $A$ expanded in the WH basis as given in Equation (\ref{eq:A_expansion}). 

\subsection{Efficient way of implementing the controlled-unitary in the qutrit framework} 

In the previous section, we saw how the unitary $U = e^{iAt}$ is implemented as a circuit. But our HHL implementation requires the controlled-$e^{iAt}$. Instead of applying control over every gate in the gadget, we propose an efficient way of doing the same for qutrits. Reference \cite{blunt_controlled_pauligadget} discusses an efficient way to decompose a controlled n-qubit gadget, into two (n+1)-qubit gadgets. In this section we extend this identity to the qutrit framework for the WH gadgets. The identity for Pauli gadgets in the qubit framework is given by - 

\begin{align*}
CU
&=
|0\rangle\!\langle 0| \otimes I
+
|1\rangle\!\langle 1| \otimes e^{i\Theta P_j} \\[4pt]
&=
e^{\,i\Theta \left( |1\rangle\!\langle 1| \otimes P_j \right)} \\[6pt]
&=
e^{\,\frac{i\Theta}{2} \left( (I - Z) \otimes P_j \right)} \\[6pt]
&=
e^{\,\frac{i\Theta}{2} \left( I \otimes P_j \right)}
\;
e^{-\frac{i\Theta}{2} \left( Z \otimes P_j \right)} .
\end{align*}
where $P_j$ is a Pauli operator acting on an n-qubit system. Consider the controlled-unitary for qutrits -
\begin{align}
U
=
|0\rangle\!\langle 0| \otimes I
+
|1\rangle\!\langle 1| \otimes e^{i\Theta W}
+
|2\rangle\!\langle 2| \otimes \left( e^{i\Theta W}\right)^{2},
\label{eq:qutrit-controlled-U}
\end{align}
where $W = c_{a,b}W_{a,b}\, + \,c^*_{a,b} W_{a,b}^\dagger$ and $W_{a,b}$ are the WH operators used in Equation (\ref{eq:A_expansion}) and their Hermitian conjugates which are added to maintain the unitarity of the exponential operators. We drop the subscripts here for convenience. Let us define an operator -
\begin{align}
G
=
|1\rangle\!\langle 1| \otimes  W
+
|2\rangle\!\langle 2| \otimes 2W. 
\label{eq:G_def}
\end{align}
Using the orthogonality of the projectors
$|0\rangle\!\langle 0|$, $|1\rangle\!\langle 1|$, and $|2\rangle\!\langle 2|$, and the fact that their powers give back the same projector,
we have -
\begin{align}
G^{n}
=
|1\rangle\!\langle 1| \otimes  W^{n}
+
|2\rangle\!\langle 2| \otimes (2W)^n .
\end{align}
Consequently, the exponential of $G$ can be evaluated via its power series expansion -
\begin{align}
e^{i\Theta G}
&=
\sum_{n=0}^{\infty} \frac{(i\Theta)^{n}}{n!} G^{n} \nonumber\\
&=
|0\rangle\!\langle 0| \otimes I
+
|1\rangle\!\langle 1| \otimes e^{i\Theta W}
+
|2\rangle\!\langle 2| \otimes e^{i\Theta 2W}.
\label{eq:qutrit-blunt-proj}
\end{align}

Now consider, 
\begin{align}
CU = e^{i\Theta G}
&=
e^{\, i\Theta
\left(
|1\rangle\!\langle 1| \otimes  W
+
|2\rangle\!\langle 2| \otimes 2 W
\right)
} , \label{eq-Blunt_step2}
\end{align}
where we need to express the projectors as a linear combination of diagonal WH operators. 

\subsubsection*{Expansion of diagonal operators in the Weyl--Heisenberg basis}

For $\omega = e^{2\pi i/3}$, we can define the Z operator as -
\begin{align}
Z = \sum_{j=0}^{2} \omega^{j} |j\rangle\!\langle j| .
\end{align}
The operators $\{I, Z, Z^{2}\}$ are diagonal in the computational basis and span the space of diagonal 1-qutrit operators.
\newtheorem{lem}{Lemma}
\begin{lem}
Any diagonal operator $D$ acting on a single qutrit can be written as
\begin{align}
D = c_{0} I + c_{1} Z + c_{2} Z^{2},
\end{align}
for suitable complex coefficients $c_{0}, c_{1}, c_{2}$.
\end{lem}

\begin{proof}
Since
\begin{align}
I &= \sum_{j=0}^{2} |j\rangle\!\langle j|, \\
Z &= \sum_{j=0}^{2} \omega^{j} |j\rangle\!\langle j|, \\
Z^{2} &= \sum_{j=0}^{2} \omega^{2j} |j\rangle\!\langle j|,
\end{align}
the operators $\{I,Z,Z^{2}\}$ are linearly independent and form a basis for the
three-dimensional space of diagonal operators.
\end{proof}

We now explicitly determine the expansion of the computational-basis projectors.

\begin{lem}
For $i \in \{0,1,2\}$, the rank-one projector $|i\rangle\!\langle i|$ admits the
expansion
\begin{align}
|i\rangle\!\langle i|
=
\frac{1}{3}
\left(
I + \omega^{-i} Z + \omega^{-2i} Z^{2}
\right).
\end{align}
\end{lem}

\begin{proof}
Assume
\begin{align}
|i\rangle\!\langle i| = c_{0} I + c_{1} Z + c_{2} Z^{2}.
\end{align}
Substituting the diagonal forms of $I$, $Z$, and $Z^{2}$ yields
\begin{align}
|i\rangle\!\langle i|
=
\sum_{j=0}^{2}
\left(
c_{0} + c_{1} \omega^{j} + c_{2} \omega^{2j}
\right)
|j\rangle\!\langle j|.
\end{align}
Equating coefficients gives the linear system
\begin{align}
\delta_{ij}
=
c_{0} + c_{1} \omega^{j} + c_{2} \omega^{2j},
\qquad j=0,1,2.
\end{align}
In matrix form,
\begin{align}
\begin{pmatrix}
\delta_{i0} \\
\delta_{i1} \\
\delta_{i2}
\end{pmatrix}
=
\begin{pmatrix}
1 & 1 & 1 \\
1 & \omega & \omega^{2} \\
1 & \omega^{2} & \omega
\end{pmatrix}
\begin{pmatrix}
c_{0} \\
c_{1} \\
c_{2}
\end{pmatrix}.
\end{align}
The coefficient matrix is the discrete Fourier transform over $\mathbb{Z}_{3}$.
Its inverse yields
\begin{align}
\begin{pmatrix}
c_{0} \\
c_{1} \\
c_{2}
\end{pmatrix}
=
\frac{1}{3}
\begin{pmatrix}
1 & 1 & 1 \\
1 & \omega^{2} & \omega \\
1 & \omega & \omega^{2}
\end{pmatrix}
\begin{pmatrix}
\delta_{i0} \\
\delta_{i1} \\
\delta_{i2}
\end{pmatrix},
\end{align}
from which the stated expression follows.
\end{proof} 

From Lemma~2, each computational-basis projector admits the expansion
\begin{align}
|i\rangle\!\langle i|
=
\frac{1}{3}
\left(
I + \omega^{-i} Z + \omega^{-2i} Z^{2}
\right),
\qquad i \in \{0,1,2\}.
\label{eq:projector-general}
\end{align}
The coefficients appearing in Equation (\ref{eq:projector-general}) follow directly
from the inverse discrete Fourier transform over $\mathbb{Z}_{3}$ and are uniquely
determined by the diagonal eigenvalues of the WH operator, $Z$.

Substituting $i=0$ into Equation (\ref{eq:projector-general}) yields
\begin{align}
|0\rangle\!\langle 0|
&=
\frac{1}{3}
\left(
I + Z + Z^{2}
\right).
\label{eq:proj0}
\end{align}

For $i=1$, using $\omega^{-1} = \omega^{2}$ and $\omega^{-2} = \omega$, we obtain
\begin{align}
|1\rangle\!\langle 1|
&=
\frac{1}{3}
\left(
I + \omega^{2} Z + \omega Z^{2}
\right).
\label{eq:proj1}
\end{align}

Finally, for $i=2$, noting that $\omega^{-2} = \omega$ and $\omega^{-4} = \omega^{2}$,
we find
\begin{align}
|2\rangle\!\langle 2|
&=
\frac{1}{3}
\left(
I + \omega Z + \omega^{2} Z^{2}
\right).
\label{eq:proj2}
\end{align}

Substituting Equations (\ref{eq:proj1}) and (\ref{eq:proj2}) in Equation (\ref{eq-Blunt_step2}) we get -

\begin{align*}
CU
&=
e^{\, i\Theta
\left(
|1\rangle\!\langle 1| \otimes  W
+
|2\rangle\!\langle 2| \otimes 2 W
\right)
} \\[4pt]
&=
e^{\, i\Theta
\left(
\frac{1}{3}
\left(
I + \omega^{2} Z + \omega Z^{2}
\right) \otimes  W
+
\frac{1}{3}
\left(
I + \omega Z + \omega^{2} Z^{2}
\right) \otimes 2 W
\right)
} \\[4pt]
&=
e^{\, \frac{i\Theta}{3}
\Bigl(
(3I + (2\omega + \omega^{2}) Z + (\omega + 2\omega^{2}) Z^{2})
\otimes  W
\Bigr)
}
\end{align*}

which can be further simplified as -
\begin{eqnarray}
CU =
\left(e^{\, i \Theta (I \otimes  W) }\right)
\left(e^{\, \frac{i\Theta (2\omega + \omega^{2}) }{3} (Z \otimes  W )} \right) \nonumber \\ 
\left(e^{\, \frac{i\Theta (\omega + 2\omega^{2}) }{3} (Z^{2} \otimes  W)}\right) . \label{eq:bluntforqutrit}
\end{eqnarray}

Thus, we obtain a product of three WH gadgets on simplification of the controlled-unitary, which is a more efficient way to implement it in a quantum circuit. The above equation is intended to be more than a formal algebraic identity; one could invoke the gate decomposition discussed in Figure \ref{fig:gadgets} to build a quantum circuit for $CU$. 

\section{Application of qutrit HHL for molecular ground-state energy calculations} \label{sec:hhllcc} 

In this section we discuss the application of qutrit HHL to quantum chemistry problems. For this purpose, we use the linearized form of coupled cluster equations as our system of linear equations - 
\begin{align}
\sum _q \langle \chi_p |H|\chi_q \rangle t_q = - \langle \chi_p | H | \Phi_0 \rangle \hspace{0.2cm},  \forall p \label{eq:lcc}
\end{align}
where, 
\begin{itemize}
    \item $|\Phi_0 \rangle$ is the Hartree-Fock determinant,
    \item $|\chi_p \rangle$ represents the $p^{th}$ excited state determinant, 
    \item $|\chi_q \rangle$ is the $q^{th}$ excited state determinant, 
    \item $H$ is the normal-ordered Hamiltonian operator, and 
    \item $t_q$ is the $q^{th}$ cluster amplitude. 
\end{itemize} 
We see that Equation (\ref{eq:lcc}) takes the form of a system of linear equations, with the vector of cluster amplitudes being the unknown vector, $\vec{x}$. 

The quantity of interest to us is the correlation energy, which is given by - 
\[
\aligned
E_{corr} &= \langle \Phi_0 | H | \chi_p \rangle t_p \\
         &= - \vec{b}^\dagger A^{-1} \vec{b} \\
         &= - k |\langle b|x\rangle|, 
\endaligned
\]
where $k = \parallel|x\rangle_{un} \parallel \; \parallel |{b}\rangle_{un} \parallel^2$. We note that we already know $|{b}\rangle_{un}$, while $\parallel|x\rangle_{un} \parallel$ can be obtained from $P(1)$ on the HHL ancilla qutrit. The molecular simulation problem is discussed in Reference \cite{adaptHHL}, one may refer to it and the references therein for detailed equations and derivations. Although we are interested in energy for the purposes of this work, one can in principle use the HHL algorithm for computing other molecular properties in the qubit or qutrit framework with an appropriate circuit module appended at the end of the HHL quantum circuit. 

The controlled-swap circuit can be appended to the HHL quantum circuit to compute the overlap of the two quantum states $|b\rangle$ and $|x\rangle$ \cite{cswap}. In the case of qubits, the absolute value of the overlap is extracted using the expression $P(0) = \frac{1}{2} \left( 1 + |\langle b|x\rangle|^2\right)$. On the other hand, for qutrit controlled-swap test, we allow the controlled-swap gate to swap the state only when it is controlled on $\ket{2}$, we can obtain the absolute value of the overlap as $P(0) = \frac{1}{9} \left( 5 + 4|\langle b|x\rangle|^2\right)$.  We add that Reference \cite{adaptHHL} uses a destructive version of the controlled-swap test, while we use the traditional version of the circuit. 

\section{Results} \label{sec:results}

\subsection{Testing with toy matrices} \label{subsec:toy}

We use the Berkeley Quantum Synthesis Toolkit (BQSKit 1.2.0) software development kit \cite{bqskit} to develop our qutrit HHL codes and to perform the simulations, since it supports computation with higher-dimensional qudits. We also develop our qubit HHL codes in the same framework so that we can then compare the results from qubit and qutrit HHLs on an equal footing. In order to test our implementation, we carry out qutrit HHL calculations on toy matrices. We begin with a simple $3 \cross 3$ diagonal matrix $A$, and $\vec{b}$ given by 
\begin{eqnarray}
 A = \begin{pmatrix}
     0.2 & 0 & 0 \\
     0 & 0.5 & 0 \\
     0 & 0 & 0.8
     \end{pmatrix}   \mathrm{and} \;
     \vec{b} = \frac{1}{\sqrt{3}}\begin{pmatrix}
    1 \\ 1\\ 1
\end{pmatrix} \label{eq:toydiag}
\end{eqnarray}
respectively. We note that our choice for $\vec{b}$ is normalized for qutrit states, and can directly be amplitude encoded as $|b\rangle = \frac{1}{\sqrt{3}}(|0\rangle + |1\rangle + |2\rangle) $. This state can be achieved using a single qutrit Hadamard gate. The statevector results obtained for this calculation using qutrit HHL for different values of $n_r$ are given in Table \ref{tab:toy matrix}. We see that our result for $\vec{x}$ is in good agreement with the solution vector obtained classically when $n_r=6$.

\newcommand{\columnn}[3]{#1 &#2 &#3 \\ } 
\begin{table}[h!]
\centering
\caption{Results from executing the qutrit HHL algorithm for the toy $A$ matrix and $\vec{b}$ described in Equation (\ref{eq:toydiag}), for different numbers of QPE clock register qutrits. PFD refers to the percentage fraction difference between our result and the expected result obtained classically. }
    \label{tab:toy matrix}
\begin{tabular}{c@{\hspace{0.7cm}}c@{\hspace{0.7cm}}c@{\hspace{0.7cm}}c} 
\hline \hline 
\textbf{$n_r$} & \textbf{$\vec{x}$} & \textbf{$|\langle b|x\rangle|$} & PFD (\%) \\
\hline 
3 & [1.76, 1.16, 0.73] & 2.1051 & 23.42 \\
4 & [2.55, 1.15, 0.73] & 2.5555 & 7.09 \\
5 & [2.63, 1.15, 0.72] & 2.6056 & 5.25 \\
6 & [2.81, 1.15, 0.72] & 2.7036 & 1.69 \\
\hline \hline 
\end{tabular}
\end{table} 

We also carry out a simple calculation with another $3 \times 3$ toy matrix, but this time non-diagonal. Our choice for the $A$ matrix and $\vec{b}$ are given by 

\begin{eqnarray}
 A = \begin{pmatrix}
     0.5 & 0.1 & 0.2 \\
     0.1 & 0.6 & 0.1 \\
     0.2 & 0.1 & 0.7
     \end{pmatrix}  \mathrm{and} \;
     \vec{b} = \begin{pmatrix}    0 \\ 1\\ 0
\end{pmatrix} \label{eq:toynondiag}
\end{eqnarray}
respectively. Again, $\vec{b}$ is normalized for qutrit states, and can be directly encoded as $|b\rangle = |1\rangle $. This state can be achieved using a 1-qutrit increment gate. The statevector results obtained for this calculation using qutrit HHL for different values of $n_r$ are given in Table \ref{tab:toy matrix2}.

\begin{table}[h!]
\centering
\caption{Results from executing the qutrit HHL algorithm for the toy $A$ matrix and $\vec{b}$ described in Equation (\ref{eq:toynondiag}), for different numbers of QPE clock register qutrits. }
\begin{tabular}{c@{\hspace{0.7cm}}c@{\hspace{0.7cm}}c@{\hspace{0.7cm}}c}
\hline \hline
\textbf{$n_r$} & \textbf{$\vec{x}$} & \textbf{$|\langle b|x\rangle|$} & PFD (\%) \\
\hline
2 & [-0.25, 1.69, -0.16] & 1.69272 & 2.80 \\
3 & [-0.28, 1.71, -0.15] & 1.70506 & 2.10 \\
4 & [-0.28, 1.73, -0.16] & 1.72855 & 0.75 \\
5 & [-0.29, 1.73, -0.19] & 1.73218 & 0.54 \\
\hline \hline
\end{tabular}
    \label{tab:toy matrix2}
\end{table} 

\subsection{Quantum chemistry simulations} \label{subsec:chem}

We now move to results from our simulations for quantum chemistry. We use the GAMESS-US software \cite{GAMESS} to generate the Graphical Unitary Group Approach Configuration Interaction Hamiltonian in the Singles and Doubles approximation (GUGA-CISD). Since the $A$ matrix is obtained using linearized coupled cluster method in the singles and doubles approximation (LCCSD), it is a principal sub-matrix of the Hamiltonian matrix, where we slice away the first row and first column (which is $\vec{b}$) of the latter \cite{psiHHL}. We benchmark our results with energies obtained using the LCCSD approximation and CISD on a classical computer. 

\subsubsection*{Case 1: 1-qutrit input state}

We consider the $H_2$ molecule in the $6-31G$ basis with different bond lengths (9 data points between 1.2 and 1.6 Bohr) to obtain a potential energy curve (PEC). For each geometry considered, the system includes four molecular orbitals that comprise one occupied and three virtuals. The outermost virtual molecular orbital is cut-off, hence our calculations consist of only three molecular orbitals for our active space that equals six spin orbitals, thus yielding a $3 \times 3$ $A$ matrix in the GUGA-CISD framework. On the HHL front, the 1-qutrit isometry circuit for each bond length is a rotation gate, $R_{02}(\theta)$, with the choice of rotation angle depending on the bond length. Table \ref{tab:isometry_thetas} of Appendix presents the values chosen for $\theta$. 

The PEC is presented in Figure \ref{fig:pec}, and the accompanying data is presented in Table \ref{apptab:pec} of the Appendix. Throughout, $n_r$ is set to 5 for our qutrit HHL calculations, while for the qubit HHL computations, we set $n_r=8$ to capture the results with comparable precision. However, for completeness, we also present the PEC with $n_r=5$ for qubit HHL in the figure as well as the table. We also add that many works that present quantum chemistry results in the quantum computing framework quote their methods' energies relative to a benchmark value, but such a presentation does not provide complete context on how well an approach has captured correlation effects, given that one almost always starts with a reference wave function like the HF state to build the more accurate one and compute energies. Therefore, we also present the HF energies in the PEC as well as in the table of results, except that in Figure \ref{fig:pec} we shift the HF energy at each bond length value, $R$, by a constant number for visual ease of discerning the total energy values provided using qubit HHL, qutrit HHL, LCCSD, and CISD methods. Our results show that as expected, qutrit HHL-LCCSD computations are able to capture correlation effects reasonably well, and the resulting total energies are at most to within $0.02$ percent of the classical LCCSD value. 

\begin{figure}[h] 
    \includegraphics[width=1.0\linewidth]{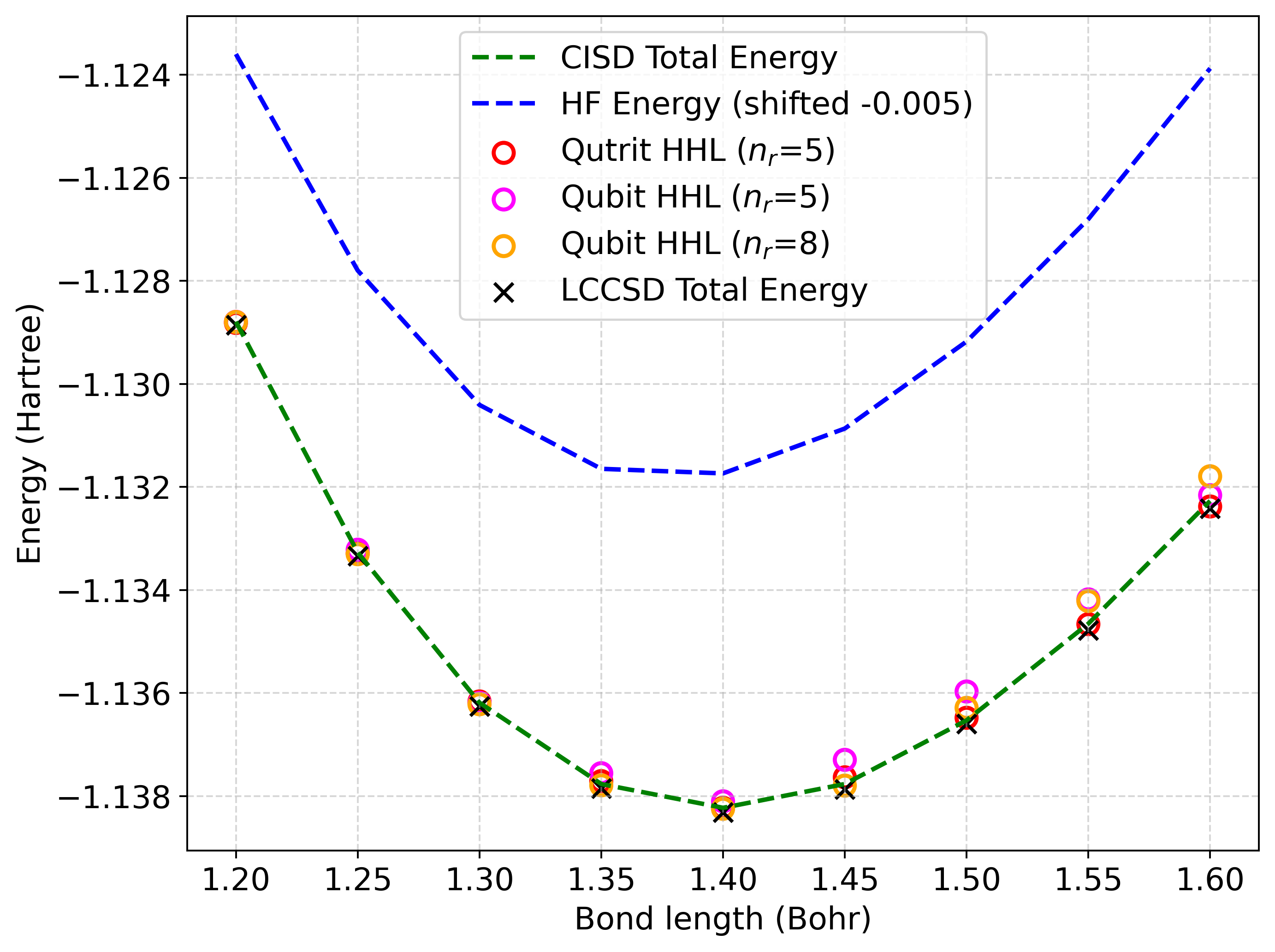}
    \caption{Potential energy curves with energies specified in units of Hartree for the $H_2$ molecule. The HF energies have each been shifted by an amount of $-0.005$ Hartree for ease of visual representation. } 
    \label{fig:pec}
\end{figure} 

We now focus on the equilibrium geometry obtained from our PEC (1.4 Bohr), to study the effect of $n_r$ on the correlation energy using qubit and qutrit HHL implementations. The results are provided in Figure \ref{fig:Ecorr_qubit_qutrit}, with Table \ref{tab:energy_qubit_qutrit} of the Appendix presenting the accompanying data. We stop at $n_r=6$ in the figure due to computational limitations from qutrit HHL. We note that for the purposes of our statevector calculations, setting $n_r=6$ would result in $2^6=64$ multi-controlled rotation gates in the controlled-rotation module in the case of qubits, whereas for qutrits, we end up with $3^6=729$ multi-controlled rotation gates. This observation underscores the need to develop more computationally efficient classical computing routines for qudit HHL implementations in general, and we defer the exercise for a future study. We now turn our attention back to the results in Figure \ref{fig:Ecorr_qubit_qutrit}, which show that for a fixed number of qudits in the clock register, qutrit HHL performs much better than its qubit counterpart. This is not surprising, given that the scaled eigenvalues are captured to $3^{n_r}$ precision for qutrits, whereas it is $2^{n_r}$ for qubits. We also see that `convergence' to the expected correlation energy value is achieved faster for qutrit HHL than qubit HHL for the same reason. We recall that the HHL algorithm involves a constant $\mathcal{C}$ (introduced in Section \ref{sec:qutritHHL}). We have modified this constant by taking its ternary expansion (for qutrits) and its binary expansion (for qubits) up to the same precision as is achieved by the number of qudits in the clock register. Figure \ref{fig:Ecorr_qubit_qutrit} shows that the correlation energies have improved with the ternary and binary expanded $\mathcal{C}$ values for qubit and qutrit HHL respectively. 

\begin{figure}[h]
    %\centering
    \hspace{-0.75cm}
    \includegraphics[width=1.0\linewidth]{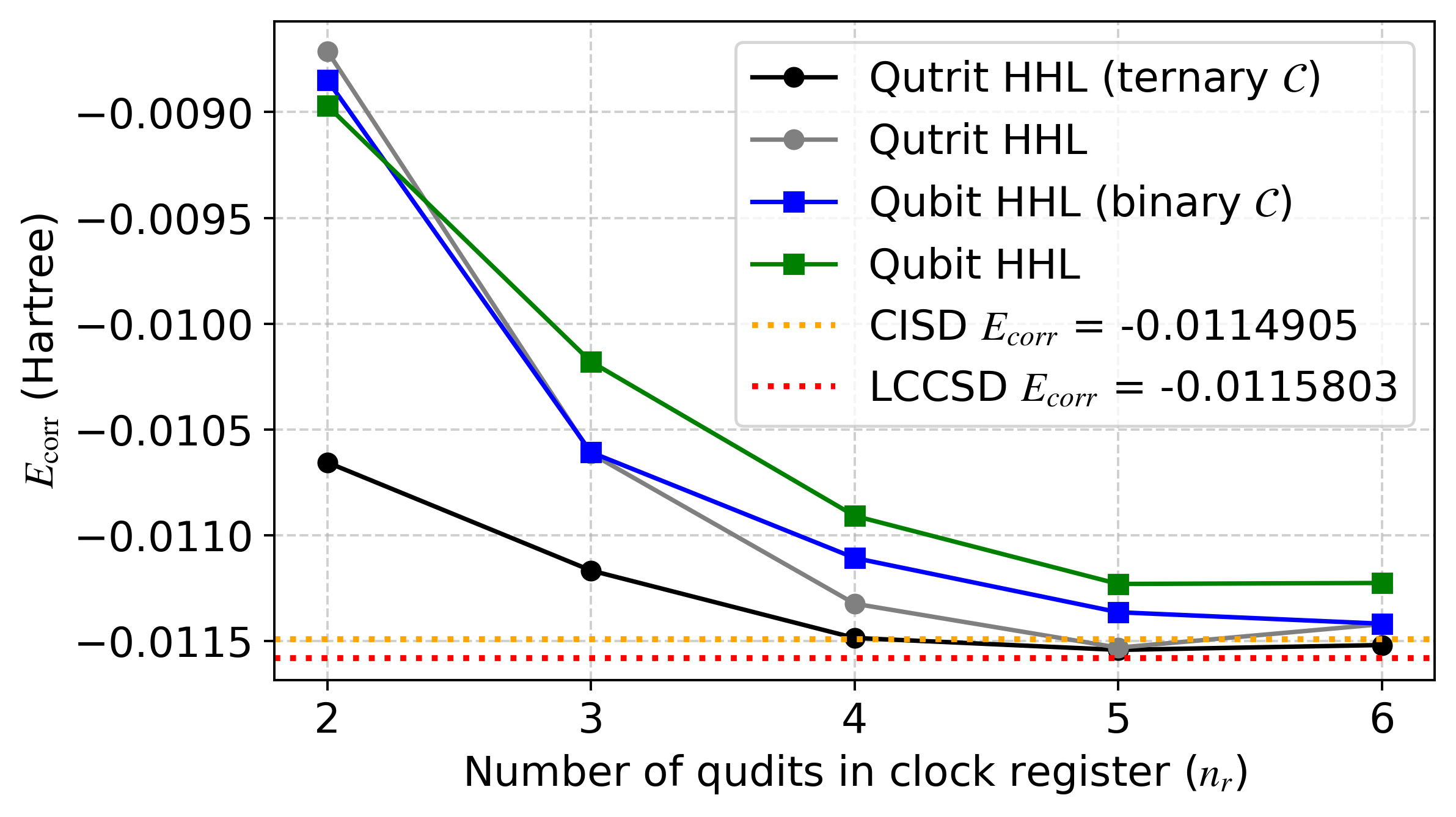}
    \caption{Correlation energy in Hartree for the $H_2$ molecule with equilibrium bond length, obtained from statevector simulations of the qubit and qutrit HHL algorithms with different values for $n_r$. } 
    \label{fig:Ecorr_qubit_qutrit}
\end{figure} 

\subsubsection*{Case 2: 2-qutrit input state}
We also perform a simple statevector calculation for the case of two qutrits in the state register following our approach described in Section \ref{sec:decomposition}. Here, we consider the same problem of the $H_2$ molecule in the $6-31G$ basis, but this time without freezing any of the orbitals, thereby working in an active space of one occupied and three virtual molecular orbitals, which makes it a total of eight spin orbitals. Using the GUGA-CISD approach, we obtain a Hamiltonian matrix of dimension $6 \times 6$, which on slicing according to the LCCSD method reduces to a $5 \times 5$ $A$ matrix, which in turn is padded to $9 \times 9$ and used as the input in our qutrit HHL implementation. For the purposes of this work, we do a proof-of-principle demonstration of the decomposition schemes only for the equilibrium geometry of $H_2$ and obtain results within $0.01$ percent precision as compared to the expected correlation energy from classical LCCSD calculations. The correlation energy obtained from qutrit HHL calculation with 5 clock registers is $-0.025401$ Hartree, while the LCCSD reference value is $-0.025527$ Hartree. Next, we move to an analysis of qubit and qutrit HHL implementations in terms of resources. 

\section{Comparison of qubit and qutrit HHLs} \label{sec:scaling}

In this section, we begin by analyzing the number of qubits and qutrits required for a given precision, followed by gate counts between qubit and qutrit HHL cases, and finally gate counts were we to consider physical Hamiltonians, employ Trotterization, and subsequently use Pauli and WH gadgets for qubits and qutrits respectively. Throughout this comparison, we do not include the cost of input state preparation as well as that of matrix loading. We also assume that the qubit-based and qutrit-based quantum computers involved have all-to-all connectivity. Furthermore, in estimating CX counts, we use decomposition schemes from References \cite{shende_decompo} and \cite{jiang_qutrit_decompo}, whereas we use the decomposition scheme discussed in Section \ref{sec:decomposition} for the purposes of application of HHL to the LCC problem. Lastly, all the qudit and gate counts are at the logical level, that is, we do not consider a fault tolerant implementation to carry out this analysis. A thorough resource estimation exercise is deferred for a future study. 

\subsection{Analyzing the number of qudits} \label{subsec:numqubits} 

\begin{enumerate}
    \item Clock register:\\ Let the required precision in decimal numbers be $p$ digits. This means that we want the error to be $\epsilon = 10^{-p}$. In the previous sections, the number of qudits in this register are denoted by $n_r$. Since the captured scaled eigenvalues $|\phi \rangle$ are stored in these $n_r$ qudits as $| \Tilde{\phi} \rangle$, $|\phi - \Tilde{\phi}| = \frac{1}{d^{n_r}}$, where $d$ is the dimension of the qudit. In such a case, the maximum error $\epsilon$ will be $\frac{1}{d^{n_r}}$, and hence 
    \[
    \aligned
    10^{-p} = \epsilon = \frac{1}{d^{n_r}} \\
    d^{n_r} = 10^p \\
    n_r = p\; log_d(10). 
    \endaligned
    \]
    Henceforth, we shall use $n_r^{(t)}$ for qutrits and $n_r^{(b)}$ for qubits.
    For qubits, $d=2$, and hence 
    \begin{align}
         n_r^{(b)}  = p\; log_2(10), \label{eq-nr-qubits}
    \end{align}
   
    while for qutrits, $d=3$, and therefore 
    
 \begin{align}
     n_r^{(t)} &= p\; log_3(10) \nonumber \\
    n_r^{(t)} &= p\; \frac{log_2(10)}{log_2(3)} \nonumber \\
     n_r^{(t)} &= \frac{n_r^{(b)}}{log_2(3)}. \label{eq-nr-qutrits}
 \end{align}

    Thus, the number of qutrits required in the clock register is lesser by a factor of $log_2(3)$ as compared to qubits. That is -
    \begin{align}
         n_r^{(t)} = \frac{n_r^{(b)}}{log_2(3)} \approx 0.63 \;n_r^{(b)}. \label{eq-nr-percent}
    \end{align} 
    For qudits in general, the reduction is by a factor of $log_2(d)$ as compared to qubits.

    \item State register: \\ In order to encode the $N \cross 1$ dimensional $\vec{b}$ in the state register, $m$ qudits are required such that $N \leq d^m$. Since we work in the binary (for qubits) and the ternary (for qutrits) frameworks, the size of $|b\rangle$ is accordingly $2^{m^{(b)}}$ or $3^{m^{(t)}}$. Here, $m^{(b)}$ and $m^{(t)}$ represent the number of qubits and qutrits in the state register respectively. The $\vec{b}$ must be encoded in these quantum states, and hence ${m^{(b)}}$ or ${m^{(t)}}$ are chosen such that they accommodate the $N \cross 1$ dimensional $\vec{b}$, while the remaining $d^m - N$ entries are taken as zeroes. From this condition, it can be shown that 
    \[
    m \geq log_d(N), 
    \]
   and the relation between the number of qubits and qutrits follows from this as
   \begin{align}
   m^{(t)} \approx \frac{m^{(b)}}{log_2(3)} \approx 0.63 m^{(b)}. \label{eq-mb_mt_relation}
   \end{align}
    Once again, we see a suppression in the number of qutrits in the state register by a factor of $log_2(3)$. This is a general expression, but for our quantum chemistry application, we can arrive at estimates for the number of qubits and qutrits in terms of the number of spin orbitals, $N_s$. For an LCCSD computation, the number of possible single and double excitations scale at most as ${N_s}^4$, which is the length of the original $\vec{b}$. For encoding it in a qubit or qutrit space, we need to pad these with zeroes until we reach a size of $2^{m^{(b)}}$(for qubits) or $3^{m^{(t)}}$(for qutrits). Table \ref{tab:state_reg qudit count} shows how the state register size grows with increasing $N_s$. We immediately see that for a calculation involving 20 spin orbitals, we need 18 qubits on the state register, whereas in a qutrit HHL implementation, we need 13. 

\begin{table}[h]
\centering
\caption{A comparison of the number of qubits and qutrits required in the state register for the LCCSD problem. $N_s$ is the number of spin orbitals resulting in ${N_s}^4$ possible excitations. For a given precision, the number of qubits and qutrits required are $m^{(b)}$ and $m^{(t)}$ respectively. }
\begin{tabular}{cccccc}
\hline \hline
$N_s$ & ${N_s}^4$ & $2^{m^{(b)}}$ & $m^{(b)}$ & $3^{m^{(t)}}$ & $m^{(t)}$ \\
\hline
2  & 16        & 16         & 4   & 27        & 3 \\
4  & 256       & 256        & 8   & 729       & 6 \\
6  & 1296      & 2048       & 11  & 2187      & 7 \\
8  & 4096      & 4096       & 12  & 6561      & 8 \\
10 & 10000     & 16384      & 14  & 19683     & 9 \\
12 & 20736     & 32768      & 15  & 59049     & 10 \\
14 & 38416     & 65536      & 16  & 177147    & 11 \\
16 & 65536     & 65536      & 16  & 177147    & 11 \\
18 & 104976    & 131072     & 17  & 531441    & 12 \\
20 & 160000    & 262144     & 18  & 1594323   & 13 \\
\hline \hline
\end{tabular}

\label{tab:state_reg qudit count}
\end{table}

\begin{figure}[h]
    \centering
    \includegraphics[width=0.9\linewidth]{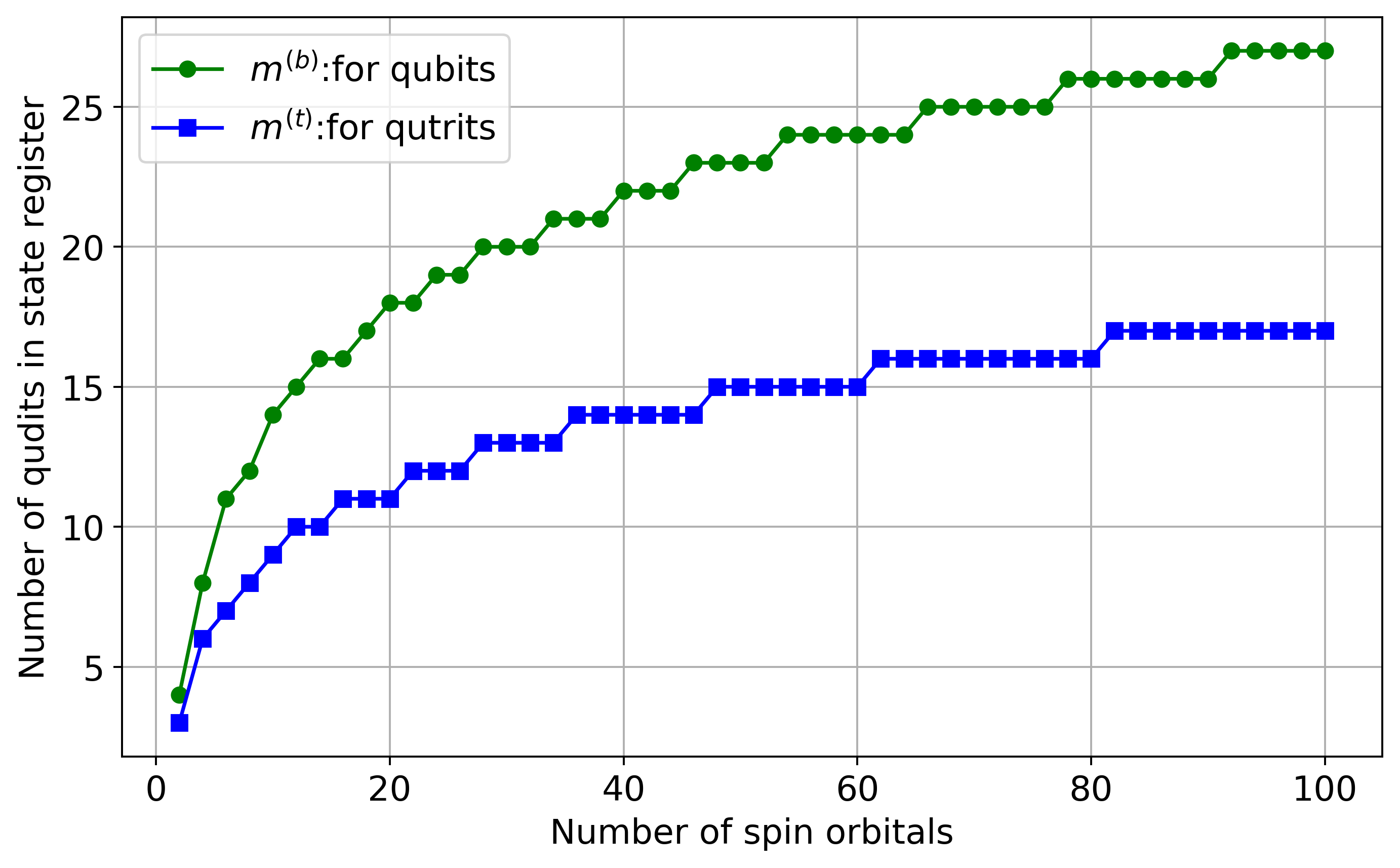}
    \caption{A comparison of the number of qubits and qutrits required in the state register for the HHL-LCCSD problem. }
    \label{fig:state_reg qudit comparison}
\end{figure}

\begin{figure}[h]
    \centering
    \includegraphics[width=0.9\linewidth]{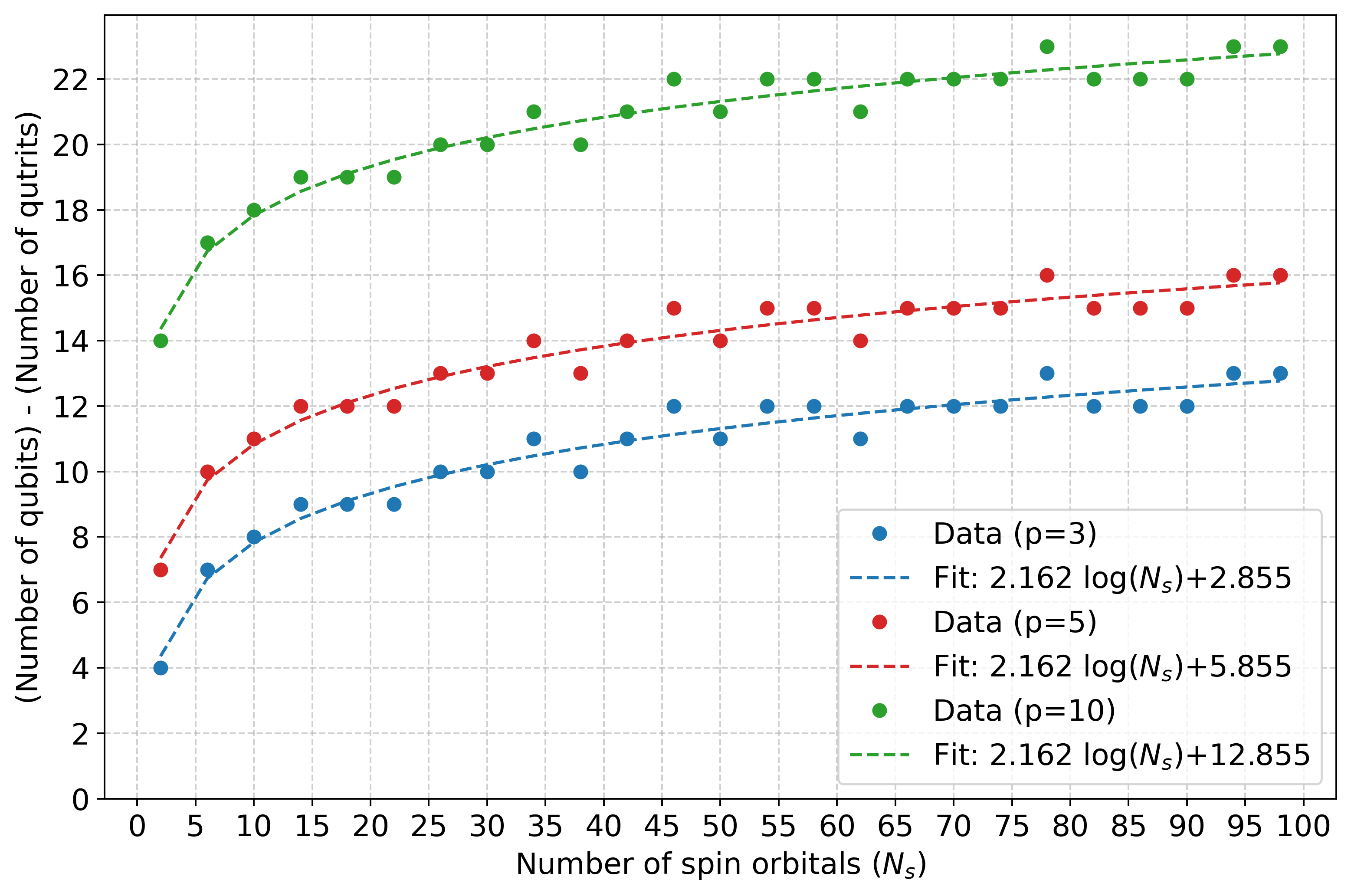}
    \caption{Curves showing the difference in the number of qubits and qutrits required for a given precision $p$, specific for the HHL-LCCSD problem. } 
    \label{fig:diff_curvefit}
\end{figure} 

 Figure \ref{fig:state_reg qudit comparison} gives a graphical representation of a similar comparison up to $N_s=100$. 

    \item HHL ancilla: \\ A single qudit is required for the HHL ancilla regardless of it being a qubit or a qutrit. 
\end{enumerate} 

Thus, in the qutrit HHL implementation,

\[ 
\mathrm{Total \; number \; of \; qutrits} \approx p \frac{log_2(10)}{log_2(3)} + \frac{log_2(N)}{log_2(3)} + 1, 
\]
as compared to the qubit HHL implementation, where
\[
\mathrm{Total \;  number \; of \; qubits} \approx p\; log_2(10) + log_2(N) + 1. 
\]

Figure \ref{fig:diff_curvefit} gives a visual representation of the difference in the total number of qutrits and qubits required for the HHL implementations with a given precision. We have done curve-fitting for three different values of precision. It can be seen from the plot that the number of qutrits required is lesser than qubits by a logarithmic amount, and that the difference increases with a demand for better precision. We add that since one needs to extract a feature of $\ket{x}$ post-HHL, which in our case is the overlap $|\langle b|x \rangle|$, the appended controlled-swap circuit module requires preparing an additional $\ket{b}$, and thus fewer qutrits than qubits as discussed earlier in the state register qudit count part. 

\subsection{Analyzing the number of 2-qudit gates} \label{subsec:numgates}

The overall resource cost can be estimated by the number of 2-qudit gates, since they are considered to be the major sources of error at the physical level. We do the counting module-wise as follows:
 
\begin{enumerate}
 
    \item Controlled-unitaries within QPE: \\
    The total number of controlled-unitaries for qubits is  $2^{{n_r}^{(b)}}- 1$, while for qutrits it is $\frac{1}{2}(3^{{n_r}^{(t)}}- 1)$. However, for a given precision $p$, using Equation (\ref{eq-nr-qutrits}), we can show that 
\begin{align}
    3^{{n_r}^{(t)}}   &= 3^{\frac{{n_r}^{(b)}}{log_2(3)}} \nonumber \\
           &= 3^{{{n_r}^{(b)}}\; log_3(2)} \nonumber  \\
           &= (3^{log_3(2)})^{{n_r}^{(b)}} \nonumber  \\
           &= 2^{{n_r}^{(b)}}. \label{eq-orders}
\end{align} 
Therefore, the number of controlled-unitaries is halved for the qutrit case. We now assume a recursive decomposition scheme for the $m^{(b)}$ qubits in the state register as proposed in Reference \cite{shende_decompo}, and a similar decomposition scheme for the $m^{(t)}$ qutrits \cite{jiang_qutrit_decompo}, and work out both the cases to find that the number of 2-qudit gates are of the same order. This is achieved because of the fact that for a given precision $m^{(b)}$ and $m^{(t)}$ are related using Equation (\ref{eq-mb_mt_relation}). We note that while the decomposition for qubits goes as $4^{m^{(b)}}$ and the one for qutrits scales as $9^{m^{(t)}}$, the number of 2-qudit gates in a controlled-unitary gate scales quadratically in the $A$ matrix size due to amplitude encoding in both the cases. 
    \item IQFT module within QPE:\\ 
    The count for 2-qudit gates in this module is $\frac{{{n_r}^{(b)}}^2 -1}{2}$ for qubits and $\frac{{{n_r}^{(t)}}^2 -1}{2}$ for qutrits, which can be simplified using Equation (\ref{eq-nr-percent}) as $ \approx \frac{ 0.39 \; {{n_r}^{(b)}}^2 -1}{2}$. Therefore, the IQFT module with qutrits has $\approx 39\%$ gates as its qubit counterpart. 
    \item Controlled-rotation module: \\ The number of multi-controlled rotation gates for the controlled rotation module is $d^{n_r}$. This means that qutrit HHL requires $3^{{n_r}^{(t)}}$ multi-controlled rotations and qubit HHL requires $2^{{n_r}^{(b)}}$ multi-controlled rotations. Using Equation (\ref{eq-orders}), we can show that the number of multi-controlled-rotation gates in the controlled-rotation module is equal for both the qubit and qutrit HHL implementations. The decomposition to 2-qudit gates again goes as in the first point. Hence, the number of 2-qudit gates in the controlled-rotation module are comparable for both the implementations. 
\end{enumerate}

\subsection{Gate counts for physical Hamiltonians and with Trotterization and Weyl-Heisenberg gadgets} \label{sec:gates_phy_H}

The gate count arising from the $e^{iAt}$ module assumed a general $A$. However, in practice, we often consider physical Hamiltonians that are often sparse, such as those that occur in atomic or molecular physics/ quantum chemistry, nuclear structure, and condensed matter calculations. In such cases, the number of Hamiltonian terms scale polynomially in the number of state register qubits. Furthermore, it is of interest to estimate gate counts when we assume Trotterization and apply our WH gadgets to the matrix exponential. Just as in the previous sub-section, we neglect the cost of preparing $\ket{b}$. We focus on the gate count from QPE, noting that the gate count from other modules such as controlled-rotation and feature extraction would be less than that from QPE. 

In the previous section, we discussed the number of controlled-unitaries being $2^{{n_r}^{(b)}}- 1$ and $\frac{1}{2}(2^{{n_r}^{(b)}}- 1)$ (Equation (\ref{eq-orders})) for qubit and qutrit HHL implementations respectively. We now estimate the gate count for each such controlled-unitary. As discussed in Section \ref{sec:decomposition}, each term in the expansion of $e^{iAt}$ upon Trotterization (first order and first step) translates to a Pauli gadget (for qubits) and WH gadget (for qutrits). For qubits, the decomposition for the controlled-unitary yields $2$ Pauli gadgets each containing $2(m^{(b)}+1)$ 2-qubit gates. Using Equation (\ref{eq:bluntforqutrit}), each controlled WH gadget is decomposed to $3$ WH gadgets each having $2(m^{(t)}+1)$  2-qutrit gates. Comparing the cases, and considering the relation between $m^{(b)}$ and $m^{(t)}$ from Equation (\ref{eq-mb_mt_relation}), we infer that the number of 2-qubit gates goes as $4(m^{(b)}+1).(poly(m^{(b)}))$, whereas the number of 2-qutrit gate count is $\frac{6}{2}(m^{(t)}+1).(Poly(m^{(t)}))$, which can be expressed in terms of $m{(b)}$ as $3(0.63m^{(b)}+1).(Poly(m^{(t)}))$. We use `Poly' for the qutrit case as opposed to `poly' for the qubit case to distinguish between the two polynomial functions. Therefore, the number of gates between the qubit and qutrit cases are still comparable upon applying HHL to physical Hamiltonians and using Trotterization and relevant gadgets for decomposing matrix exponentials. 

\section{Conclusion} \label{sec:conclusion} 

In this work, we extend the HHL algorithm from its qubit form to the qutrit framework, which includes proposing a simplified controlled-rotation module and designing the circuits. For the latter, we come up with Weyl-Heisenberg (WH) gadgets beyond the versions involving only $Z$ and $Z^2$ strings from literature. We also design and derive in detail a circuit for decomposing the controlled-unitaries in terms of these WH gadgets. We then develop qubit and qutrit HHL codes in the open-source BQSKit framework from scratch. To demonstrate the working of the qutrit HHL algorithm, we carry out toy matrix calculations as well as quantum chemical calculations involving 1-qutrit state registers to generate the potential energy curve of the model $H_2$ molecule. Furthermore, we make use of our gadgets to find the ground state energy in the equilibrium geometry of the same system, but with a 2-qutrit state register. We obtain the total energies to within 0.02 percent precision for the 1-qutrit state example and to within 0.01 percent for the 2-qutrit state example (the equilibrium geometry case), and also benchmark our results with CISD and LCCSD calculations using classical methods. 

We then carry out an analysis of the qudit count and the gate count between qubit and qutrit HHL implementations, all at the logical level, that is, we omit fault tolerance. We report the following observations in contrast with the qubits case: we show that for a given precision, the required number of clock register qutrits is lower by a factor of log$_2(3)$ relative to qubit HHL; it goes as log$_2(d)$ for qudit HHL in general. To amplitude encode the same amount of information from the $N \times 1$ vector, $\vec{b}$, into the HHL input state, we observe that number of qutrits required is $\lceil \mathrm{log}_3(N)\rceil$ while for qubits it is $\lceil \mathrm{log}_2(N)\rceil $. With regard to the number of 2-qudit gates, the requirement for the QFT module within  QPE is reduced to 39\% of the number of 2-qubit gates needed in qubit HHL, whereas the number of controlled unitaries in the QPE module reduces to 50\% of those necessary for qubit HHL. The number of multi-controlled rotation gates in the controlled-rotation module is found to be the same in both cases. However, when we decompose each controlled-unitary, and multi-controlled rotation gate in terms of 2-qudit gates, we find that the 2-qudit gate count for qubit and qutrit HHL is comparable (under the assumption that the decomposition schemes are those from Refs. \cite{shende_decompo} and \cite{jiang_qutrit_decompo}). We also find that upon assuming sparse physical Hamiltonians, where the decomposition of $A$ in the Pauli or the Weyl-Heisenberg basis yields only a polynomial number of terms in the number of qudits, and upon further applying Trotterization and Pauli or Weyl-Heisenberg gadgets, the number of gates from the QPE module are still comparable for qubit and qutrit HHLs. It is also worth adding that the analysis excludes state preparation and matrix loading costs, and assumes that the underlying qubit and qutrit hardwares have all-to-all connectivities. 

In summary, the scope of our work involves introducing qutrit HHL and designing gadgets to realize controlled-$e^{iAt}$ for a practical implementation, as well as develop codes from scratch and show proof-of-concept computations of ground state energies in a model molecular system. Furthermore, we compare qubit and qutrit HHL implementations at a logical level by estimating qudit counts and 2-qudit gate counts. We conclude that if one is able to realize quantum hardware with good quality qutrits, a qutrit HHL implementation is relatively cheaper in the number of qudits, while the number of gates is comparable for both qubit and qutrit HHL implementations. 

\textit{Future directions:} while the next logical step is a thorough resource estimation for a molecule of choice, such an exercise is deferred for a future study as it would involve qutrit hardware-based details such as gate times, gate errors, connectivity, and qubit transport, as well as qutrit QEC considerations. Likewise, while a study on the effect of different types of noise on molecular energies computed via qubit and qutrit HHL algorithms is an interesting follow-up step to the elements of our current work, this too is deferred to a future study. Lastly, it would also be interesting to study the performance of qubit and qutrit HHL approaches on real quantum hardware. However, that too is postponed for future as executing a task on noisy hardware can involve extensive compilation steps, including but not limited to circuit optimization and transpilation, as well as possibility of improving results with error mitigation. 

\section*{Acknowledgments} 
VSP acknowledges support from CRG grant (CRG/2023/002558). TP and VSP thank Dr. Akshaya Jayashankar for her constant support with conceptual ideas throughout the duration of the work. TP thanks Ms. Aashna Zade for her help with GAMESS, and Mr. Peniel Bertrand Tsemo for discussions on quantum chemistry. TP and VSP thank Dr. Ed Younis for his support with BQSKit, and thank their group members for reading the manuscript and giving their comments (Aashna, Peniel, Disha, Fazil, Akshaya, Smriti and Kalam). 

%\clearpage 

\begin{table*}[t]
\centering
\Large \textbf{APPENDIX}
\end{table*}

\begin{appendices} 

\appendix
\renewcommand{\thesection}{A.\arabic{section}}
\renewcommand{\thefigure}{A.\arabic{figure}}
\renewcommand{\thetable}{A.\arabic{table}}
\renewcommand{\theequation}{A.\arabic{equation}} 
\setcounter{section}{0}
\setcounter{figure}{0}
\setcounter{equation}{0} 
\setcounter{table}{0} 

\begin{table*}[h]
\centering
\caption{$\theta$ (in radians and rounded off to four decimal places) used to implement the isometry module in qutrit HHL for every bond length ($R$) in Bohr, for the PECs  in Figure~\ref{fig:pec}. }
\begin{tabular}{c|@{\hspace{0.5cm}}c@{\hspace{0.3cm}}c@{\hspace{0.3cm}}c@{\hspace{0.3cm}}c@{\hspace{0.3cm}}c@{\hspace{0.3cm}}c@{\hspace{0.3cm}}c@{\hspace{0.3cm}}c@{\hspace{0.3cm}}c} 
\hline \hline 
$R$ &1.20&1.25&1.30&1.35&1.40&1.45&1.50&1.55&1.60 \\   \hline 
$\theta$ &1.0296&1.0074&0.9845&0.9615&0.9383&0.9152&0.8920&0.8690&0.8465 \\ 
\hline \hline 
\end{tabular}
    \label{tab:isometry_thetas}
\end{table*}

\begin{table*}[t]
\centering
\caption{Table showing the data for all the relevant energy values used for generating the PECs in Figure \ref{fig:pec}. All the energies are specified in units of Hartree and bond lengths are in units of Bohr. All the energies are rounded
off to the sixth decimal place. }
\begin{tabular}{c|ccc@{\hspace{0.75cm}}c@{\hspace{0.5cm}}c@{\hspace{0.5cm}}c}
\hline \hline 
$R$ & $E_{\mathrm{HF}}$ & $E_{\mathrm{CISD}}$ & $E_{\mathrm{LCCSD}}$ & $E_{\mathrm{qutrit\ HHL}}$ with $n_r=5$ & $E_{\mathrm{qubit\ HHL}}$ with $n_r=5$ & $E_{\mathrm{qubit\ HHL}}$ with $n_r=8$ \\
&  & ($E_{\mathrm{CISD}}^{\mathrm{corr}}$) & ($E_{\mathrm{LCCSD}}^{\mathrm{corr}}$) & ($E_{\mathrm{qutrit\ HHL}}^{\mathrm{corr}}$) & ($E_{\mathrm{qubit\ HHL}}^{\mathrm{corr}}$) & ($E_{\mathrm{qubit\ HHL}}^{\mathrm{corr}}$)   \\ \hline 
1.20 & -1.118598 & -1.128792 & -1.128855 & -1.128816 & -1.128795 & -1.128798 \\
 &  & (-0.010194) & (-0.010257) & (-0.010218) & (-0.010197) & (-0.010199) \\ \hline 
1.25 & -1.122798 & -1.133270 & -1.133338 & -1.133287 & -1.133218 & -1.133305 \\ 
 &  & (-0.010472) & (-0.010540) & (-0.010489) & (-0.010420) & (-0.010507) \\ \hline 
1.30 & -1.125408 & -1.136187 & -1.136261 & -1.136163 & -1.136192 & -1.136217 \\
 &  & (-0.010779) & (-0.010853) & (-0.010755) & (-0.010785) & (-0.010809) \\ \hline 
1.35 & -1.126649 & -1.137766 & -1.137847 & -1.137709 & -1.137556 & -1.137791 \\
 &  & (-0.011117) & (-0.011198) & (-0.011061) & (-0.010907) & (-0.011143) \\ \hline 
1.40 & -1.126737 & -1.138228 & -1.138317 & -1.138227 & -1.138101 & -1.138246 \\
 &  & (-0.011491) & (-0.011580) & (-0.011490) & (-0.011364) & (-0.011509) \\ \hline 
1.45 & -1.125866 & -1.137766 & -1.137865 & -1.137636 & -1.137297 & -1.137797 \\
 &  & (-0.011900) & (-0.011999) & (-0.011770) & (-0.011431) & (-0.011930) \\ \hline 
1.50 & -1.124177 & -1.136529 & -1.136601 & -1.136480 & -1.135970 & -1.136288 \\
 &  & (-0.012352) & (-0.012424) & (-0.012303) & (-0.011794) & (-0.012111) \\ \hline 
1.55 & -1.121802 & -1.134653 & -1.134780 & -1.134663 & -1.134182 & -1.134214 \\
 &  & (-0.012851) & (-0.012978) & (-0.012861) & (-0.012380) & (-0.012412) \\ \hline 
1.60 & -1.118877 & -1.132271 & -1.132414 & -1.132379 & -1.132160 & -1.131793 \\
 &  & (-0.013394) & (-0.013537) & (-0.013502) & (-0.013283) & (-0.012917) \\ \hline 
\end{tabular}

\label{apptab:pec}
\end{table*} 

\begin{table*}[h]
    \centering
    \caption{Correlation energies in units of Hartree obtained with different values of $n_r$ (the number of clock register qudits) for the $H_2$ molecule with equilibrium bond length, from statevector simulations of the qubit and qutrit HHL implementations. The LCCSD and the CISD values for the correlation energies are provided for reference. All the energies are rounded off to the sixth decimal place. $\mathcal{C}$ is the minimum eigenvalue of $A$ after scaling, while binary and ternary $\mathcal{C}$ are its binary and ternary expansions to a precision of $n_r$ qubits and qutrits respectively. }
\begin{tabular}{c@{\hspace{0.9cm}}c@{\hspace{0.3cm}}c@{\hspace{0.9cm}}c@{\hspace{0.3cm}}c}
    \hline \hline
    \textbf{$n_r$} &
    \multicolumn{2}{c@{\hspace{0.8cm}}}{\textbf{$E_{corr}^{(b)}$}} &
    \multicolumn{2}{c}{\textbf{$E_{corr}^{(t)}$}} \\ 
    \cline{2-3} \cline{4-5}
     & {With C} & {With binary C} & {With C} & {With ternary C} \\ \hline
    2 & -0.008970 & -0.008848 & -0.008712 & -0.010656   \\
    3 & -0.010180 & -0.010608 & -0.010613 & -0.011167  \\ 
    4 & -0.010908 & -0.011107 & -0.011323 & -0.011485   \\ 
    5 & -0.011230 & -0.011364 & -0.011531 & -0.011542  \\ 
    6 & -0.011225 & -0.011418 & -0.011420 & -0.011519  \\ 
     \hline 
     LCCSD & \multicolumn{4}{c}{-0.011580}  \\ 
     CISD & \multicolumn{4}{c}{-0.011490}  \\ \hline \hline
\end{tabular}
\label{tab:energy_qubit_qutrit}
\end{table*}

\end{appendices}

\clearpage

% \bibliography{sample}

\bibliographystyle{apsrev4-2}
\bibliography{sample2}

\end{document}